\documentclass[journal]{IEEEtran}
\def\ps@headings{%
\def\@oddhead{\mbox{}\scriptsize\rightmark \hfil \thepage}%
\def\@evenhead{\scriptsize\thepage \hfil \leftmark\mbox{}}%
\def\@oddfoot{}%
\def\@evenfoot{}}
\makeatother
\usepackage{booktabs} 
\usepackage{url}
\usepackage{svg}
\usepackage{etoolbox}
\usepackage{cuted}
\usepackage{lipsum}
\usepackage{amssymb}
\usepackage{array}
\usepackage{amsmath}
\usepackage{tikz}
\usepackage{color}
\usepackage{dirtytalk}
\usepackage{algpseudocode}
\usepackage{graphicx}
\usepackage{mdframed}
\usepackage{tcolorbox}
\usepackage{comment}
\usepackage{cite}
\usepackage{amsmath,amssymb,amsfonts}
\usepackage{textcomp}
\usepackage{url}
\usepackage{etoolbox}
\usepackage{cuted}
\usepackage{lipsum}
\usepackage{array}
\usepackage{color}
\usepackage{dirtytalk}
\usepackage{algpseudocode}
\usepackage[english]{babel}
\usepackage[utf8x]{inputenc}
\usepackage{amsmath,amsthm}
\usepackage{multirow}
\usepackage[ruled,vlined]{algorithm2e}
\usepackage{fancyvrb}
\usepackage{pifont}
\usepackage{booktabs,makecell}

\usepackage{siunitx}
\usepackage{ragged2e}

\theoremstyle{definition}

\theoremstyle{plain}

\ifCLASSINFOpdf
\else
\fi
\begin{document}

\title{Aggregated Zero-knowledge Proof and Blockchain-Empowered Authentication for Autonomous Truck Platooning}

\author{Wanxin Li,~\IEEEmembership{Member, IEEE}, Collin Meese,~\IEEEmembership{Student Member, IEEE}, \\  Hao Guo*,~\IEEEmembership{Member, IEEE}, and Mark Nejad,~\IEEEmembership{Senior Member, IEEE}

\thanks{Manuscript received January 8, 2022; revised October 11, 2022; accepted April 10, 2023. \textit{(Corresponding author: Hao Guo.)}}

\IEEEcompsocitemizethanks{
\IEEEcompsocthanksitem Wanxin Li is with the Department of Communications and Networking, Xi'an Jiaotong-Liverpool University, Suzhou, 215123, China (e-mail: wanxin.li@xjtlu.edu.cn).
\IEEEcompsocthanksitem Collin Meese and Mark Nejad are with the Department of Civil and Environmental Engineering, University of Delaware, Newark, Delaware, 19716, USA (e-mail: \{cmeese, nejad\}@udel.edu).
\IEEEcompsocthanksitem Hao Guo is with the School of Software, Northwestern Polytechnical University, and the Yangtze River Delta Research Institute of NPU, Taicang, 215400, China (e-mail: haoguo@nwpu.edu.cn).
}}

\markboth{IEEE Transactions on Intelligent Transportation Systems,~Vol.~XX, No.~XX,~2023}%
{Shell \MakeLowercase{\textit{et al.}}: Bare Advanced Demo of IEEEtran.cls for IEEE Computer Society Journals}

\IEEEtitleabstractindextext{%
\begin{abstract}
\justifying
Platooning technologies enable trucks to drive cooperatively and automatically, providing benefits including less fuel consumption, greater road capacity, and safety. To establish trust during dynamic platooning formation, ensure vehicular data integrity, and guard platoons against potential attackers in mixed fleet environments, verifying any given vehicle's identity information before granting it access to join a platoon is pivotal. Besides, due to privacy concerns, truck owners may be reluctant to disclose private vehicular information, which can reveal their business data to untrusted third parties. To address these issues, this is the first study to propose an aggregated zero-knowledge proof and blockchain-empowered system for privacy-preserving identity verification in truck platooning. We provide the correctness proof and the security analysis of our proposed authentication scheme, highlighting its increased security and fast performance. The platooning formation procedure is re-designed to seamlessly incorporate the proposed authentication scheme, including the \emph{1st catch-up} and \emph{cooperative driving} steps. The blockchain performs the role of verifier within the authentication scheme and stores platooning records on its digital ledger to guarantee data immutability and integrity. In addition, the proposed programmable access control policies enable truck companies to define who is allowed to access their platoon records. We implement the proposed system and perform extensive experiments on the Hyperledger platform. The results show that the blockchain can provide low latency and high throughput, the aggregated approach can offer a constant verification time of 500 milliseconds regardless of the number of proofs, and the platooning formation only takes seconds under different strategies. The experimental results demonstrate the feasibility of our design for use in real-world truck platooning.




\end{abstract}

}

\maketitle
\pagestyle{headings} 

\IEEEdisplaynontitleabstractindextext

%
\IEEEpeerreviewmaketitle

\ifCLASSOPTIONcompsoc
\IEEEraisesectionheading
{\section{Introduction}\label{sec:introduction}}
\else
\section{Introduction}
\label{sec:introduction}
\fi

\IEEEPARstart
{T}{ruck} platooning, enabled by vehicular ad hoc network (VANET), is expected to affect the freight industry profoundly. The application involves connecting two or more trucks in a convoy with wireless connectivity and vehicle automation. By sharing control parameters (e.g., speed, direction, acceleration) amongst the vehicles, platoon members can achieve cooperative automated driving that results in less fuel consumption, greater roadway capacity, and, most importantly, safer operation. More specifically, a platoon member can safely follow its preceding vehicle at a much shorter headway (e.g., 0.6 seconds) than a human driver's conventional two-second following headway. The drastic headway reduction could yield a lane capacity of 4,250 vehicles per hour\cite{van2006impact}, double the existing lane capacity. Fuel efficiency is also expected to improve as a by-product of the short following headway. According to the Japan ITS Energy project, 13\% of fuel can be saved with a 10 m intra-platoon following gap at 80 km/h \cite{tsugawa2014results}. Truck platooning also allows the driver to disengage from driving tasks. Human error was estimated to be responsible for 94\% of traffic accidents in the U.S. \cite{nhtsa2015Critical}. Compared to human drivers, automated driving systems could achieve a much shorter response time and more accurately assess dynamic traffic conditions. 

There are two models for truck platooning: the road train model and the opportunistic model\footnote{https://logisticsviewpoints.com/2019/07/22/there-are-two-models-for-truck-platooning-which-will-win/}. The former involves platooning multiple trucks having the same origins and destinations. The latter aligns more with the VANET structure, where trucks from 
mixed fleets would need a mechanism to identify platooning opportunities and fairly appropriate platooning roles. To take full advantage of the estimated 45 thousand platoon-able miles in the U.S.\footnote{https://www.fleetowner.com/technology/article/21702545/for-truck-platooning-to-work-heres-what-has-to-happen}, as an example, an opportunistic model is necessary to accommodate heterogeneous fleets and serve the \$732 billion U.S. trucking market \footnote{https://www.trucking.org/economics-and-industry-data}. However, it is still unclear how to integrate the opportunistic model under the context of a \emph{mixed fleet} environment, where a platoon encompasses vehicles from different companies operating on distinctly heterogeneous and isolated systems. Mixed fleet platooning was demonstrated successfully in the Grand Cooperative Driving Challenge (GCDC) \cite{ploeg2012introduction}, but security and privacy were not a concern due to the nature of the competition. To the best of our knowledge, this is the first study to propose a privacy-preserving and efficient authentication scheme for truck platooning formation with mixed fleets where trucks are owned and operated by different companies.

The security of the platooning systems, which protects them from attacks and unauthorized access to proprietary information about a specific vehicle or fleet specifications, under mixed fleet scenarios has not been 
studied thus far, though awareness of such aspects has gained increasing attention\footnote{https://www.truckinginfo.com/355628/vtti-to-study-autonomous-trucks-in-mixed-fleets}\cite{chen2020smart, singh2020integrating}. To ensure the security of the system, it is crucial to be able to verify a given vehicle's identity information prior to granting it access to join a platoon, ensuring the platoon's integrity and guarding against potential spoofing attacks. At the same time, truck owners may be reluctant to disclose private vehicular information to an untrusted third party and other participants due to privacy concerns. For example, the potential for revealing an enterprise's intricate logistics operations through analysis of its truck platooning data by competitors would discourage participation in the system. Consequently, hiding vehicles' identity information is essential for adopting dynamic platooning in mixed fleet networks. Moreover, if the dynamic platoon formation service is controlled by a centralized entity, the system becomes vulnerable and can be at risk for data breaches and other attacks which can expose private user data. As a result, safeguarding the privacy of truck owners while simultaneously providing a means of dynamic identifier verification for other heterogeneous trucks presents a key challenge to realizing the benefits of truck platooning systems in a mixed fleet environment. 

In relation to dynamic truck platooning, blockchain presents some desirable properties for creating a robust, dynamic, and decentralized authentication system. As a fault tolerant and decentralized network technology, blockchain has become a hot research topic since its initial deployment in the Bitcoin protocol for use as a distributed digital currency ledger \cite{nakamoto2008bitcoin}. Lately, many research studies have highlighted the applicability of blockchain technology to a wide array of subject areas outside of cryptocurrency, such as healthcare \cite{ guo2022hybrid, 9259231, guo2019access}, smart cities \cite{chen2020blockchain, li2019blockchain,  guo2022hierarchical} and intelligent transportation systems (ITS) \cite{9290445, meese2022bfrt,   shaygan2022traffic}. Specifically, there are two distinct forms of blockchain technology: permissionless and permissioned. Both forms of blockchain technology inherently provide immutability properties, network reliability, and data provenance, making them potential candidates for supporting a dynamic truck platooning system. 

However, permissionless blockchain systems are entirely transparent-by-design, presenting privacy challenges when the proposed system is processing and recording sensitive user data. For example, in Bitcoin, the entirety of the ledger is publicly visible and accessible to all participants without the flexibility for defining permission levels or access controls directly on the chain. Additionally, the consensus mechanisms in permissionless blockchains (e.g., proof of work) allow for open participation, are generally token-centric, and computations are designed to be artificially complex in an effort to protect the network from attackers. These properties present stark inefficiencies, unnecessarily high bandwidth and energy consumption, as well as environmental concerns.  

Conversely, a permissioned blockchain system supports the addition of permission levels and access control mechanisms for protecting on-chain data, providing a framework capable of supporting a system where data ownership and privacy are required. However, these properties of permissioned blockchain only allow for the protection of the sensitive user data once it has been committed to the ledger and do not preserve privacy during the validation process. To remedy this issue, some researchers have employed cryptographic schemes, such as variations of the zero-knowledge proof (ZKP), in creative ways to protect the data prior to the storage process \cite{partala2020non, li2020blockchain, li2021p}. ZKP provides a method for one party (the prover) to convince another entity (the verifier), without revealing any information other than the fact they possess the said knowledge \cite{goldwasser1989knowledge}.

\subsection{Contributions and Organization}

In this paper, we study a privacy-preserving and efficient authentication paradigm for autonomous truck platooning in mixed fleets, which has not been investigated in previous work. More specifically, we made the following contributions to autonomous trucking platooning formation:

\begin{itemize}
    
    \item We proposed an aggregated and efficient zero-knowledge authentication scheme for privacy-preserving identity verification atop a permissioned blockchain network. The blockchain performs the role of verifier within the authentication scheme and stores platooning records on its digital ledger to guarantee data immutability and integrity. Correspondingly, we re-designed the platoon formation procedure and integrated the proposed authentication scheme within the formation process.

    \item The proposed aggregated zero-knowledge proof can verify the identity \textit{without revealing any private information}. We provide both the correctness proof and security analysis of the proposed authentication scheme. In contrast to a single-proof design, the aggregated proof provides increased security while offering constant verification time regardless of the number of proofs.

    \item Using Hyperledger platform, we designed and prototyped the permissioned blockchain network and conducted extensive performance benchmark testing. To protect the privacy of stored platooning records, we design programmable access control policies that enable the data owners to define what entities can access their data while recording all retrieval events in an immutable access log. 
    
    \item We developed the aggregated zero-knowledge proof with the Hyperledger Ursa cryptographic library. Experiments are conducted to quantify the end-to-end running time of our proposed authentication scheme with respect to the number of proofs. The results demonstrate that our proposed scheme provides low-latency and privacy-preserving authentication for mixed fleet truck platooning.

\end{itemize}

In the context of truck platooning, our proposed extensions to ZKP verify and protect a vehicle's sensitive identifying information during the verification process before storing a record on the blockchain ledger. In our design, the number of proofs required for successfully authenticating a truck equals the number of registered companies in the mixed fleet network. This decentralized authentication process establishes trust among participating companies. By proposing to incorporate multiple proofs into the verification process, we provide better security for the system versus a single-proof approach due to the increased difficulty required to forge multiple proofs simultaneously. Additionally, the aggregation process provides faster verification performance compared to sequential verification. The proposed aggregated zero-knowledge proof unifies multiple zero-knowledge proofs into one aggregated proof, such that the validity of the aggregated proof implies the validity of all individual proofs. As a result, our authentication method can provide efficient identity verification performance regardless of the number of individual proofs.

The remainder of the paper is structured as follows: Section 2 outlines the related work and the background knowledge; Section 3 gives the overview and design goals of the proposed system; the detailed system building blocks are provided in Section 4; the correctness and security analysis are presented in Section 5; in Section 6, we conduct the implementation and provide thorough experimental results for evaluation; lastly, we conclude the work in Section 7.

\section{Related Work and Preliminaries}
In this section, we first provide a comparative analysis of the closely related works as illustrated in Table \ref{tab:comparison}. Next, we outline the important preliminary knowledge of truck platooning, zero-knowledge proof and permissioned blockchain in the remaining subsections. 

\begin{table*}[t]
\centering
\caption{Comparative Analysis of the Proposed Scheme with the Existing Schemes}
\label{tab:comparison}
\scalebox{0.87}{
\begin{tabular}{|c|c|c|c|c|}
\hline
\textbf{References} & \textbf{Authentication Use Case}                           & \textbf{Blockchain Type} &\textbf{Privacy-preserving Method} & \textbf{Implementation} \\ \hline
Lin et al.\cite{lin2020bcppa}   & Communication in VANET                  & Permissionless  & PKI-based signature                & Ethereum       \\ \hline
Feng et al.\cite{feng2019bpas}   & General VANET & Permissioned & Attribute-based encryption (ABE) & Hyperledger Fabric                    \\ \hline
Yao et al.\cite{yao2019bla}   & Vehicular fog service          & N/A             & N/A                       & N/A            \\ \hline
Kaur et al.\cite{kaur2019blockchain}    & Vehicular fog computing          & N/A             & Elliptic curve cryptography (ECC)                       & N/A            \\ \hline
Liu et al.\cite{liu2020blockchain}    & Vehicular edge computing       & N/A             & N/A                       & Python         \\ \hline
Hexmoor et al.\cite{hexmoor2018blockchain}    & Information sharing in platoon & N/A             & N/A                       & N/A            \\ \hline
Ying et al.\cite{ying2019bavpm}    & Record management for platoon   & Permissionless  & ECDSA signature           & Ethereum       \\ \hline
Ying et al.\cite{ying2020beht}    & Toll charging for platoon      & Permissionless  & Aggregated signature      & Ethereum       \\ \hline
Chen et al.\cite{chen2019smart}    & Payment for platoon service    & N/A             & N/A                       & N/A            \\ \hline
Proposed scheme & Platooning formation in mixed fleets & Permissioned & Aggregated ZKP + Access Control  & Hyperledger Fabric + Hyperledger Ursa \\ \hline
\end{tabular}
}
\end{table*}

\subsection{Related Work}
In regard to transportation, blockchain has been proposed as a way to decentralize emerging intelligent transportation systems, by establishing a secured, trusted, and decentralized ecosystem that can better use existing ITS resources. More specifically, building a secure and trustable information-sharing framework for vehicle-to-vehicle (V2V) and vehicle-to-infrastructure (V2I) communications has been a key research focus over the past few years \cite{mollah2020blockchain}. For example, the authors of \cite{9217515} propose a blockchain-based event recording system in the context of dynamic autonomous vehicle environments. In \cite{8010820}, a blockchain-based dynamic key management system for heterogeneous networks is proposed with a focus on ensuring security for vehicular communication systems within the ITS. Li et al. propose a decentralized traffic management system to protect data integrity and privacy in a scenario of multiple blockchain-based connected vehicular networks \cite{9210529, li2019blockchain}. 

However, given the dynamic and real-time nature of vehicular networks, ensuring identity authentication while preserving privacy has become a challenge. Since 2018, some researchers have explored using blockchain-based design as a solution to these problems. Lin et al. proposed the blockchain-based conditional privacy-preserving authentication (BCPPA) verification protocol based on PKI signature and Ethereum blockchain to facilitate secure communication in VANETs \cite{lin2020bcppa}. Nevertheless, experiments demonstrate that deploying the necessary smart contracts on the public Ethereum blockchain can be excessively costly to the user due to the gas fees paid to miners. In \cite{feng2019bpas}, Feng et al. introduced a framework called blockchain-assisted privacy-preserving authentication system (BPAS) that provides authentication automatically in VANETs and preserves vehicle privacy at the same time. Still, the proposed approach does not support aggregated proofs. Yao et al. proposed a blockchain-assisted lightweight anonymous authentication (BLA) mechanism for distributed vehicular fog services, which is provisioned to driving vehicles \cite{yao2019bla}; however, only the consensus algorithm is simulated while ignoring the cost of other blockchain maintenance operations. An effective cross-datacenter authentication and key-exchange scheme using blockchain and Elliptic curve cryptography for vehicular fog computing were presented by Kaur et al., but no blockchain experiments were conducted \cite{kaur2019blockchain}.

In vehicular edge computing, a blockchain-empowered group-authentication scheme was proposed by Liu et al. \cite{liu2020blockchain} for vehicles with decentralized identification based on secret sharing and dynamic proxy mechanism. Notably, the authors proposed a hybrid proof-of-work and proof-of-trust consensus mechanism, but proof-of-work mining results in unnecessary energy consumption for energy-conscious vehicles. To protect platoon member privacy and security while providing a rapid sharing of telemetry data, Hexmoor et al. proposed an adaptation of blockchain technology to information exchanges among vehicles traveling in a platoon \cite{hexmoor2018blockchain}. However, the \emph{cooperative driving} steps are not considered during platoon formation. Ying et al. proposed a dynamic autonomous vehicle platoon (AVP) management protocol by implementing Ethereum \cite{ying2019bavpm}. While the authors successfully designed smart contracts with a low gas cost for the user, the verification time is bottlenecked by Ethereum's slower consensus time compared to permissioned blockchain algorithms (e.g., PBFT). Later in \cite{ying2020beht}, they proposed a blockchain-based efficient highway toll paradigm for the opportunistic platoon. In addition, an aggregated signature was introduced to accelerate the authentication procedure. To improve the urban traffic condition and reduce accidents, Chen et al. proposed a platoon-driving model for autonomous vehicles in a free-flow traffic state. In this study, a smart contract is employed to enable the payment based on a blockchain between the platoon head (PH) and platoon members (PMs), avoiding malicious and false payments \cite{chen2020smart}.

From the literature study, we have observed that blockchain-empowered identity authentication in platooning formation had not been investigated in previous works. Existing blockchain-based authentication schemes were introduced for different use cases in vehicular networks. However, many existing schemes lack for: (1) privacy-preserving methods except for leveraging the decentralization properties of blockchain in authentication \cite{yao2019bla, liu2020blockchain, hexmoor2018blockchain}, and (2) concrete blockchain designs, such as blockchain type for use and implementations \cite{yao2019bla, kaur2019blockchain, liu2020blockchain, hexmoor2018blockchain, chen2019smart}. Hence, we propose an authentication scheme based on permissioned blockchain and aggregated zero-knowledge proof to resolve vehicular network security and privacy challenges, specifically for use in autonomous truck platooning formation with mixed fleets. The comparative analysis of the proposed authentication scheme with the existing schemes is described as shown in Table \ref{tab:comparison}.

\subsection{Truck Platooning}
\begin{figure}[t]
\centering
\includegraphics[width=0.45\textwidth]{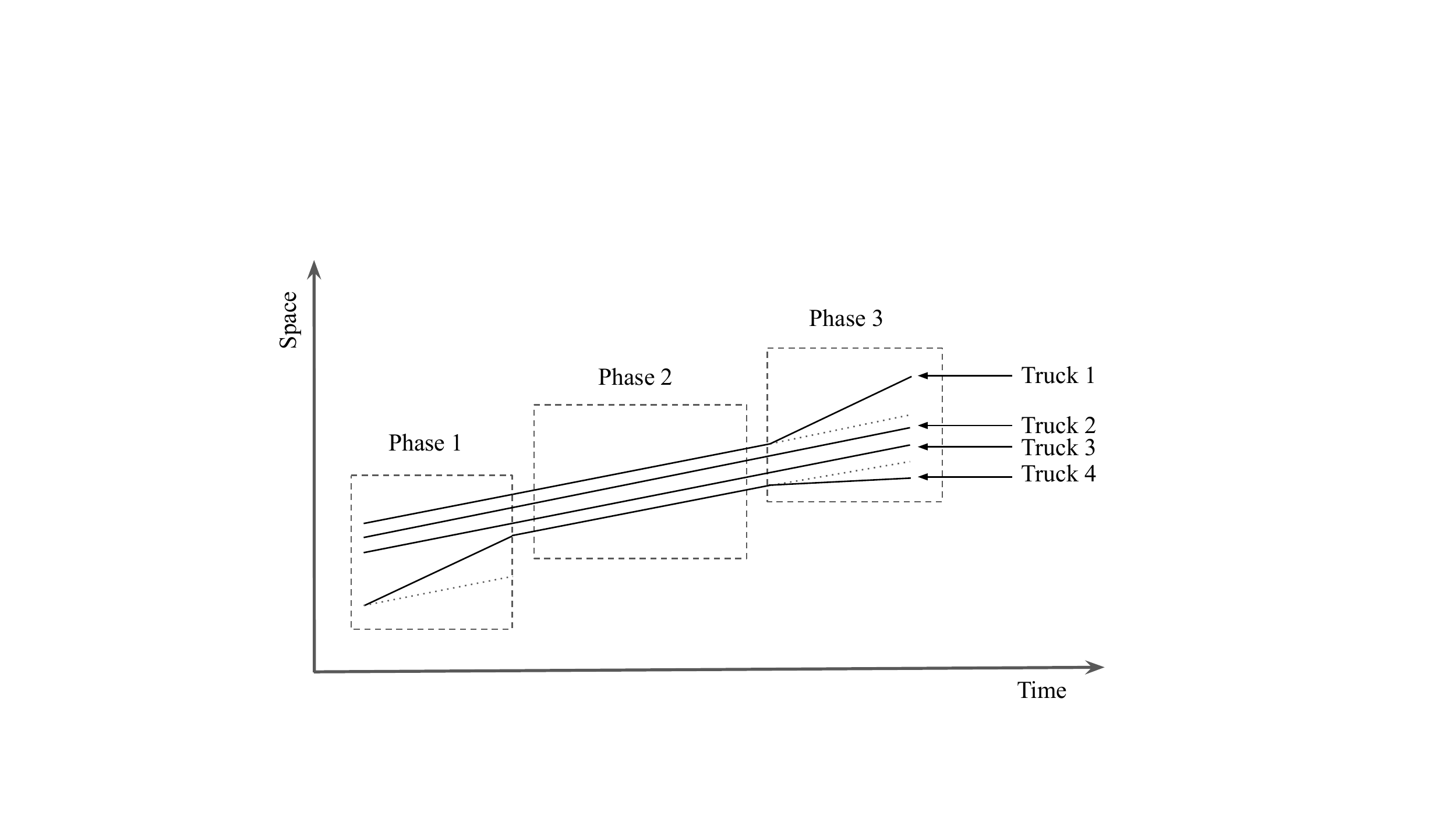}
\caption{Time-space diagram showing platooning phases: phase 1 formation, phase 2 maintaining and phase 3 separation. Each sloping line represents the trajectory of a truck in motion during the platooning phases, and there are four lines portrayed as an example. The focus of this research is on the first phase of platooning: the formation phase.} 
\label{fig:platooning phases}
\end{figure}

The growth rate of freight transportation has led to an increase in the number of trucks on highways. Other than negative environmental impacts, trucks typically have a lower speed in comparison with cars, resulting in the reduction of the highway capacity. The platooning of trucks can be considered a potential approach to mitigate the negative effects of trucks on highway traffic streams. In addition, platooning of trucks reduces the air drag forces affecting them, thus contributing to less fuel consumption and emissions \cite{bonnet2000fuel}. One would expect the benefits of truck platooning to increase as the spacing between trucks decreases. Therefore, many studies investigate the deployment of automated vehicles connected by using wireless communications for truck platooning. To form a platoon, the trucks decrease the spacing between them and adjust their speeds so that they can move together along the highway \cite{zhang2004control}. Fig. \ref{fig:platooning phases} shows the different phases of platooning. 

Platoons of trucks can be formed before the trucks leave their origin (road train model) or as they move along the highway (opportunistic model). An example of the former case is when trucks belonging to the same trucking company form a platoon before departing their depot. In the latter case, a more general scheme, the trucks, which may belong to one or various trucking companies, form a platoon along their way (Phase 1 of Fig. \ref{fig:platooning phases}). These trucks remain in the platoon for the shared portion of their paths (Phase 2 of Fig. \ref{fig:platooning phases}). Finally, a platoon separates when one or more of the trucks forming the platoon leave it or the platoon needs to be split (Phase 3 of Fig. \ref{fig:platooning phases}). This research focuses on the first phase of platooning, that is, the formation phase. \textit{The goal is to verify new joining trucks from different companies in a privacy-preserving manner, which leads to the secure formation of the platoon on the road.}

\subsection{Zero-Knowledge Proof}
The zero-knowledge proof, first proposed by Goldwasser, Micali, and Rackoff \cite{goldwasser1989knowledge}, provides a method by which a prover can convince a verifier that they know a secret message $m$, without disclosing any information apart from the fact the prover possesses $m$. The power of this proof is centered around the idea that, while it is trivial to prove one possesses a secret by simply revealing the secret, it becomes a significant challenge to prove such possession without disclosing any additional information about the secret or the secret itself. Further research in this area has produced two main categories of ZKP protocols: interactive ZKP and non-interactive ZKP. In an interactive ZKP scheme, the prover and the verifier engage in a dialogue, and the verifier is required to generate multiple challenges for the prover to solve. By solving the challenges correctly, the verifier is convinced, without a doubt, that the prover indeed possesses the knowledge. In a non-interactive scheme, the prover generates the proof directly, which can be independently verified to prove possession of the given knowledge. In consequence, it reduces the communication overhead of the proof system. A valid ZKP must satisfy the following three properties:

\begin{itemize}
    \item Completeness: If the statement is true, a prover will convince an honest verifier of this fact.
        
    \item Soundness: If the statement is false, no cheating prover can convince the honest verifier that it is true.
    
    \item Zero-knowledge: If the statement is true, no verifier learns anything other than the fact that the statement is true.
    
\end{itemize}

The inherent properties of ZKP can be leveraged to ensure input validity for blockchain transactions without revealing sensitive information during the validation process. One example of ZKP in existing blockchain systems is Zcash, which utilizes the zk-SNARK (zero-knowledge succinct non-interactive arguments of knowledge) protocol for achieving consensus on transaction information (e.g., amounts, wallet address balances, etc.) without exposing any of the actual details \cite{hopwood2016zcash}. In addition, Bulletproofs are another emerging and non-interactive method for applying ZKP to blockchain systems, offering short proofs that do not require a trusted setup. Furthermore, proof aggregation is a trending feature that can increase security and efficiency in systems where multiple proofs are involved. For example, Bulletproofs support aggregation of range proofs so that a prover can prove that $n$ commitments lie in a given range by providing an additive $O(\log(n))$ group elements over the length of a single proof \cite{bunz2018bulletproofs}.

\subsection{Permissioned Blockchain}
Blockchain comes in two primary varieties: permissionless and permissioned \cite{miller2019permissioned}. In the permissionless case (e.g., Bitcoin), membership is entirely open and anyone can join the network and view all of the transactions. In contrast, a permissioned blockchain is a closed membership network, where a consortium of one or more entities will make collaborative decisions about membership, data access controls and governance policies. In permissioned blockchain, anyone interested in validating transactions or viewing data on the network needs to get approval from a certificate authority. This is useful for companies, banks and institutions that are comfortable complying with the regulations and are very concerned about having complete control of their data. Due to the ability to control membership, permissioned blockchain systems can utilize lighter-weight consensus algorithms than their permissionless counterparts. Furthermore, programmable access controls can be defined within permissioned blockchain systems, providing fine-grained control for on-chain data. These properties make permissioned blockchain technology more attractive for certain applications requiring high transaction throughputs with low latencies. 

\section{System Overview and Design Goals}
\subsection{System Overview}
\begin{figure}[t]
\centering
\includegraphics[width=0.45\textwidth]{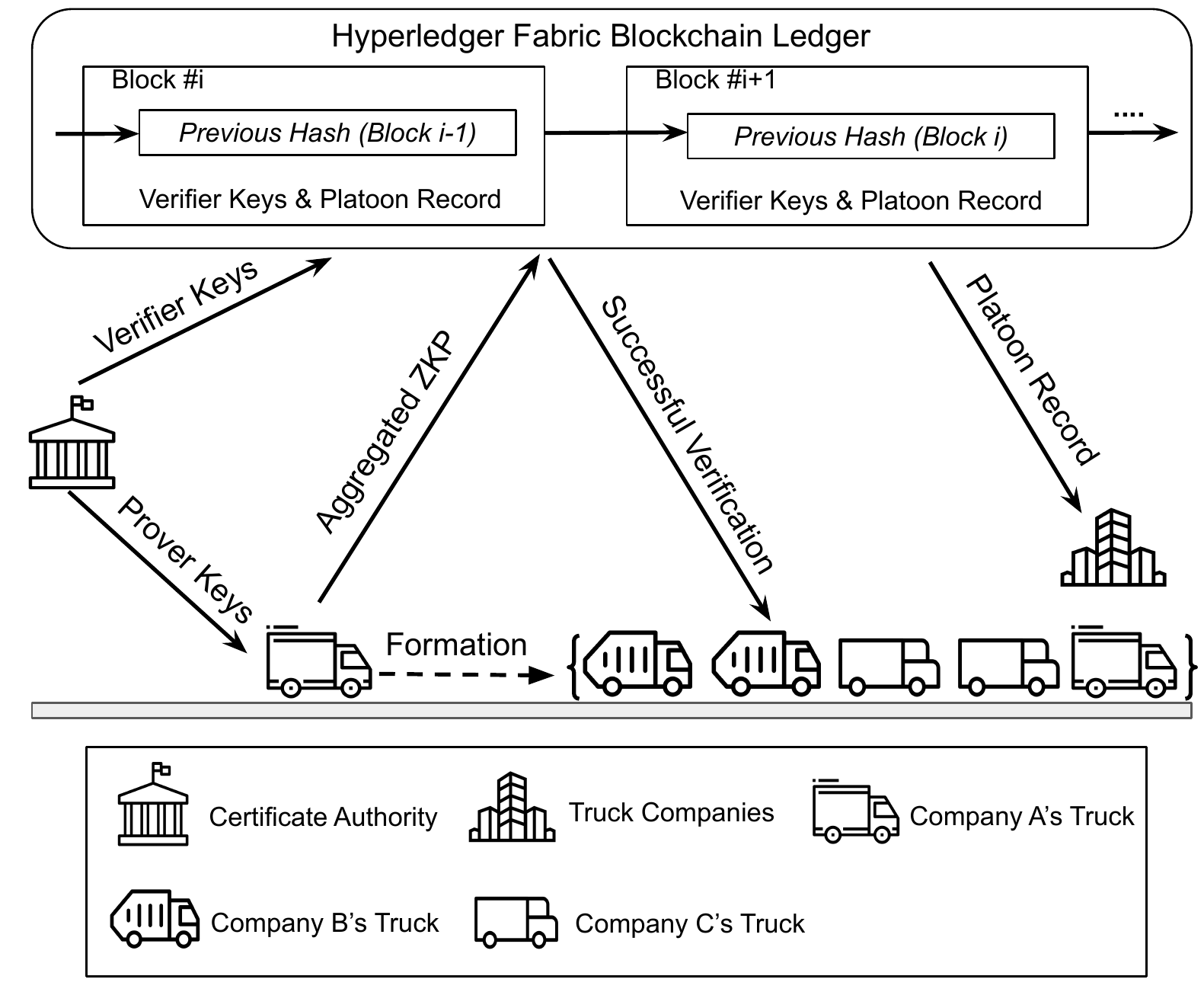}
\caption{Entities in the proposed authentication system for autonomous truck platooning in a mixed fleet network. (Note that we give three truck companies and six autonomous trucks as an example here. The proposed authentication system in this paper can serve for an arbitrary number of truck companies and autonomous trucks. )}
\label{fig:arch}
\end{figure}

By referring to Fig. \ref{fig:arch}, we first define the following entities that will take part in the proposed authentication system:

\begin{itemize}
    
    \item Permissioned Blockchain: Our permissioned blockchain is utilized as the controller of the system and serves as the verifier to validate the identities of new trucks before joining any mixed fleet platoon. It also provides a tamper-proof transaction ledger for recording verifier keys and platoon records.
    
    \item Certificate Authority (CA): A certificate authority is an entity that manages the private identifiers of autonomous vehicles (e.g., MAC address\footnote{https://en.wikipedia.org/wiki/MAC\_address}) and issues key pairs to data provers and data verifiers. In practice, an agency that already has all vehicle registration information, such as the Department of Motor Vehicles (DMV), can function as the proposed system's certificate authority.
    
    \item Trucking Companies: Trucking companies are the clients in our proposed blockchain network and manage different groups of trucks. Trucking companies may need to retrieve the platooning histories of their owned trucks for use in practical applications, such as determining the optimal platoon size on each route to reduce fuel consumption, and improve efficiency and safety.
    
    \item Autonomous Truck: An autonomous truck is a participant of mixed fleet platoons and also acts as a prover to prove its identity during the authentication process before joining any existing platoon.

\end{itemize}

\begin{figure*}[t]
\includegraphics[width=0.85\textwidth]{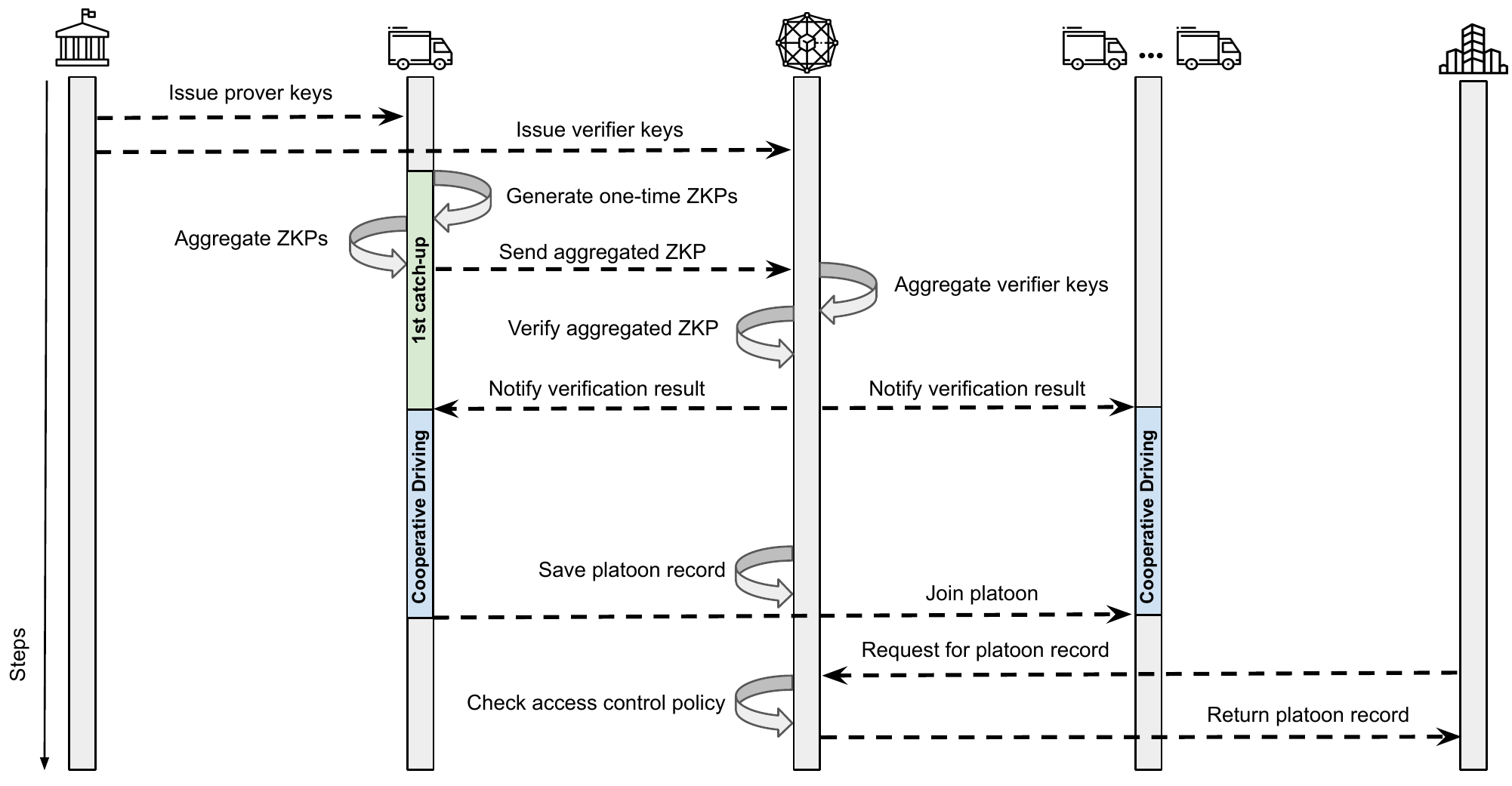}
\centering
\caption{Workflow of the proposed authentication system for autonomous truck platooning in a mixed fleet network.}
\label{fig:workflow}
\end{figure*}

Fig.~\ref{fig:workflow} illustrates the system workflow for autonomous truck platooning. In the beginning, the trucking companies register their vehicles on the blockchain network. Each individual truck is issued a set of prover keys and matching verifier keys, where the number of key pairs corresponds to the number of companies. This way, trust is built between every autonomous truck and trucking company in a shared vehicular network. Verifier keys are stored directly on the blockchain ledger and used to validate a truck's identity during the verification process. Key issuance is a one-time process performed when a vehicle is initially registered in the system. 

When entering the vehicle-to-vehicle communication range, an autonomous truck can request to join the existing platoon by first generating a set of one-time zero-knowledge proofs based on its identifier, using each of its prover keys, respectively. Next, the vehicle performs a local proof aggregation to generate a short zero-knowledge proof and sends it to the blockchain network as a transaction. This process invokes a smart contract which performs the required verifier key aggregation and subsequently commits the result to the ledger. If the verification is successful, the autonomous truck will be notified and drive cooperatively to join the platoon. Furthermore, a trucking company can submit a retrieval request to obtain platoon records from the ledger, and a smart contract will verify the request against the defined access control policy. If the policy is satisfied, the smart contract will return the requested records to the company.

\subsection{Design Goals}
From the specification of the proposed system architecture, it captures the following security goals:

\begin{itemize}
    \item Correctness: Without knowing the standalone truck's identity information, the generated proof can be correctly verified by the blockchain network during platooning formation phase.
    
    \item Maximal Privacy-preserving: An autonomous truck's sensitive and private information is only accessible to its truck company and certificate authority (e.g., DMV), and can never be disclosed to the other participants in a mixed fleet vehicular network. 
    
    \item Efficiency: The aggregated zero-knowledge proof is introduced for constructing the verification protocol. Instead of sending every single proof to the blockchain network for identity verification, the aggregated proof can provide fast and constant verification time by varying the number of proofs.
    
    \item Data Ownership: A truck company has full possession and control of its owned truck records on the shared blockchain ledger. It includes the ability to record, access and derive benefit from its owned truck's data, and the right to assign these access privileges to others by programmable access control policies.
    
\end{itemize}

\section{System Building Blocks}

This section describes the construction details of the proposed system, which includes the platooning formation procedure, the aggregated zero-knowledge proof for authentication and the permissioned blockchain network with access control policies. 

\subsection{Platooning Formation Procedure}
\label{sec: platoon formation}

Table \ref{tab:variable2} shows the variables we use in constructing the platooning formation procedure. A generalized platooning formation procedure is summarized as follows:

\begin{enumerate}
\item A standalone truck enters the communication radius of an existing platoon (e.g., 300 m in dedicated short-range communications). Vehicle-to-vehicle information is exchanged for the possibility of the standalone truck joining the existing platoon.
\item Suppose the standalone truck is determined to be a good candidate for joining the platoon. In that case, it starts to send the authentication request to the blockchain network while accelerating its speed to shorten the distance from the platoon, which is called the \emph{1st catch-up} step.
\item If the authentication is passed, \emph{cooperative driving} will start with one of the platooning strategies for platooning formation. If the authentication is failed, the \emph{cooperative driving} will not start.
\end{enumerate}

\begin{table}[ht]
    \renewcommand{\arraystretch}{1.45}
    \centering
    \caption{Variables in the Platooning Formation Procedure}
    \label{tab:variable2}
    \begin{tabular}{cc}
    \hline
      {\thead{\textbf{Variable}}} & {\textbf{Description}}\\ \hline
      $Q$ & number of trucks in the platoon\\
      $R$ & maximum vehicle-to-vehicle communication radius \\
      $\chi$ & the standalone truck\\
      ${P_i}(t)$  &  position of truck $i$ at time $t$ \\
      $P_{ij}(t)$ & relative position of trucks $i$ and $j$ at time $t$ \\
      $v_{i}(t)$  &  speed of truck $i$ at time $t$ \\
      $v_{i}^{min}$ & minimum speed for truck $i$\\
      $v_{i}^{max}$ & maximum speed for truck $i$\\
      $a_{i}(t)$ & acceleration for truck $i$ at time $t$\\
      $a_{i}^{min}$ & minimum acceleration for truck $i$\\ 
      $a_{i}^{max}$ & maximum acceleration for truck $i$\\
      $\Gamma$ & time of authentication (1st catch-up)\\
      $\Theta$ & time of cooperative driving\\
      $\mathrm{T}$ & total time of platooning formation\\
    \hline
    \end{tabular}
\end{table}

Assume a platoon $\bar{Q}=\left \{ 1, \cdots, i, j, \cdots, Q \right \}$, such as truck $j$ follows truck $i$ and truck number 1 is the leader in the platoon. The relative position of two trucks in duration $\Delta t$ is expressed in Equation \ref{eq:relativePosition}. For each vehicle, the change of position in duration $\Delta t$ is calculated using Equation \ref{eq:displacement}.

\begin{equation}
\label{eq:relativePosition}
    P_{ij}(t+\Delta t) = P_{ij}(t) + [P_{i}(t+\Delta t)-P_{j}(t+\Delta t)].
\end{equation}

\begin{equation}
\label{eq:displacement}
\begin{split}
    P_{i}(t+\Delta t)= &v_{i}(t) \Delta t + \frac{1}{2}a_{i}(t)(\Delta t)^2\\ 
    0 \leq v_{i}^{min} &\leq v_{i}(t) \leq v_{i}^{max},\\
    a_{i}^{min} \leq &a_{i}(t) \leq a_{i}^{max}.
\end{split}
\end{equation}

In the \emph{1st catch-up} step, given a standalone truck $\chi$ and the total authentication time $\Gamma$, the relative position of the platoon and the standalone truck will be shortened from $R$ to $P_{i \chi}(t+\Gamma), \forall i\in \bar{Q}$ based on Equation \ref{eq:relativePosition}, in which $P_{i \chi}(t+\Gamma)$ can be expressed in Equation \ref{eq:1st catch-up}.

\begin{equation}
\label{eq:1st catch-up}
\begin{split}
    P_{i \chi}(t+\Gamma) = R - [P_{\chi}(t+\Gamma) - P_{i}(t+\Gamma)], \forall i\in \bar{Q}.
\end{split}
\end{equation}

Once the  authentication is passed, the standalone truck $\chi$ will be added into the set of platoon $\bar{Q}$ for the next step, where the \emph{cooperative driving} begins. Fundamentally, there are three platooning strategies: 

\begin{enumerate}
\item The \emph{2nd catch-up} strategy, where the standalone truck continues to accelerate to catch up with the platoon. 
\item The \emph{slow-down} strategy, where the platoon decelerates to allow for the standalone truck to catch up. 
\item The \emph{hybrid} strategy, where the standalone truck and the platoon move cooperatively towards an intermediate state.
\end{enumerate}

With the relationship in Equations \ref{eq:relativePosition} and \ref{eq:displacement}, the kinematic equations for the duration of a \emph{cooperative driving} can be expressed in Equation \ref{eq:platoonStrategy} based on the three platooning formation strategies. 
The shortest time $\Theta$ is dictated by the vehicle pair that takes the longest time to reach the targeting platoon state. 

\begin{align}
\Theta=
\left\{\begin{matrix}
\textrm{max}\left \{ \frac{\Delta P_{1 i}}{\Delta v^{max}_{i}-\Delta v_{1i}} \right \}, \forall i\in \bar{Q} & \textrm{2nd catch-up } \\ 
\textrm{max} \left \{ \frac{\Delta P_{i Q}}{-\Delta v^{max}_{i} - \Delta v_{i Q}} \right \},  \forall i\in \bar{Q} &  \textrm{slow-down }\\ 
\textrm{max} \left \{ \frac{\Delta P_{i Q}}{\Delta v^{\textrm{max}}_{Q} - \Delta v^{\textrm{max}}_{i}-\Delta v_{i Q}} \right \}, \forall i\in \bar{Q}&  \textrm{hybrid}
\end{matrix}\right.
\label{eq:platoonStrategy}
\end{align}

Considering $\Delta P_{1 \chi} \gg \Delta P_{1i}$ and $\Delta P_{\chi Q} \gg \Delta P_{i Q}$ ($i \in \bar{Q}, \chi \neq i$) during the \emph{cooperative driving} step, Equation \ref{eq:platoonStrategy} can be simplified as:

\begin{align}
\Theta=
\left\{\begin{matrix}
\frac{\Delta P_{1 \chi}}{\Delta v^{max}_{\chi}-\Delta v_{1 \chi}}, & \textrm{2nd catch-up } \\ 
\frac{\Delta P_{\chi Q}}{-\Delta v^{max}_{\chi} - \Delta v_{\chi Q}}, &\textrm{slow-down }\\ 
\frac{\Delta P_{\chi Q}}{\Delta v^{\textrm{max}}_{Q} - \Delta v^{\textrm{max}}_{\chi}-\Delta v_{\chi Q}},& \textrm{hybrid}
\end{matrix}\right.
\label{eq:platoonStrategy2}
\end{align}

Note that $\Delta P_{1 \chi}$ and $\Delta P_{\chi Q}$ can be calculated as:

\begin{equation}
\begin{split}
    \Delta P_{1 \chi} &= [P_{1}(t) - P_{\chi}(t)] - [P_{1}(t+\Gamma) - P_{\chi}(t+\Gamma)] \\
    &= R - [v_{1}(t)\Gamma + \frac{1}{2}a_{1}(t){\Gamma}^2 - v_{\chi}(t)\Gamma - \frac{1}{2}a_{\chi}(t){\Gamma}^2]\\
    &=R - [v_{1}(t)-v_{\chi}(t)]\Gamma - \frac{1}{2}[a_{1}(t) - a_{\chi}(t)]{\Gamma}^2
\end{split}
\end{equation}

\begin{equation}
\begin{split}
    \Delta P_{\chi Q} &= [P_{\chi}(t) - P_{Q}(t)] - [P_{\chi}(t+\Gamma) - P_{Q}(t+\Gamma)] \\
    &= -R - [v_{\chi}(t)\Gamma + \frac{1}{2}a_{\chi}(t){\Gamma}^2 - v_{Q}(t)\Gamma - \frac{1}{2}a_{Q}(t){\Gamma}^2]\\
    &= -R - [v_{\chi}(t)-v_{Q}(t)]\Gamma - \frac{1}{2}[a_{\chi}(t) - a_{Q}(t)]{\Gamma}^2
\end{split}
\end{equation}

According to Fig. \ref{fig:workflow}, the end-to-end authentication time $\Gamma$ is a summation of time costs from the \emph{generate one-time ZKPs} step to the \emph{verify aggregated ZKP} step. Finally, the total time in platooning formation phase is represented as Equation \ref{eq:total}:

\begin{equation}
\label{eq:total}
\begin{split}
    \mathrm{T} &= \Gamma + \Theta \\
    &= \left\{\begin{matrix} \Gamma + \frac{R - [v_{1}(t)-v_{\chi}(t)]\Gamma + \frac{1}{2}[a_{\chi}(t) - a_{1}(t)]{\Gamma}^2}{\Delta v^{max}_{\chi}-\Delta v_{1 \chi}}, & \textrm{2nd catch-up } \\ 
    \Gamma + \frac{R - [v_{Q}(t) - v_{\chi}(t)]\Gamma + \frac{1}{2}[a_{\chi}(t) - a_{Q}(t)]{\Gamma}^2}{\Delta v^{max}_{\chi} + \Delta v_{\chi Q}}, &\textrm{slow-down }\\ 
    \Gamma + \frac{R - [v_{Q}(t) - v_{\chi}(t)]\Gamma + \frac{1}{2}[a_{\chi}(t) - a_{Q}(t)]{\Gamma}^2}{\Delta v^{\textrm{max}}_{\chi} + \Delta v_{\chi Q} -\Delta v^{\textrm{max}}_{Q}},& \textrm{hybrid}
\end{matrix}\right.
\end{split}
\end{equation}

\subsection{Aggregated Zero-Knowledge Proof}

In this subsection, we introduce the aggregated zero-knowledge proof scheme that validates the identity of autonomous trucks in an efficient and privacy-preserving manner. Table \ref{tab:variable1} shows the variables for constructing the aggregated zero-knowledge proof.

\begin{table}[ht]
    \renewcommand{\arraystretch}{1.45}
    \centering
    \caption{Variables in the Aggregated Zero-Knowledge Proof}
    \label{tab:variable1}
    \begin{tabular}{cc}
    \hline
      {\thead{\textbf{Variable}}} & {\textbf{Description}}\\ \hline

      $m$ & secret identifying message (e.g., MAC address)\\
      $n$ & number of truck companies in the mixed fleet network\\
      $\Omega_\lambda$ & one-time ZKP ($\lambda = 1, 2, \cdots, n$)\\
      $\Omega$ & aggregated one-time ZKP for $\Omega_1, \Omega_2,\cdot\cdot\cdot,\Omega_n$\\
      $sk_\lambda$ & prover key for generating the $\Omega_\lambda$\\
      $pk_\lambda$ & verifier key for verifying the $\Omega_\lambda$\\
      $pk$ & aggregated verifier key for verifying the $\Omega$\\    
      $h$ & one-time hash digest of the secret message $m$\\
      $r$ & result for verifying the aggregated ZKP $\Omega$\\
      $\mathbb{G}$ & multiplicative cyclic group of prime order $p$\\
      $g$ & generator in $\mathbb{G}$\\
      $e$ & elliptic curve for bilinear pairing\\
    \hline
    \end{tabular}
\end{table}

We then describe how to construct the aggregated zero-knowledge proof verification scheme in the following five algorithms:

\textit{$KeyGen(~) \longrightarrow (sk_\lambda, pk_\lambda)$}: This algorithm chooses a random $sk_\lambda \in \mathbb{Z}_p$ and computes $pk_\lambda = g^{sk_\lambda} \in \mathbb{G}$. Return the prover key $= sk_\lambda$ and the verifier key $= pk_\lambda$ for $\lambda = 1,..., n$.

\begin{algorithm}[h]
\label{alg: zkp-keygen}
\SetAlgoLined
\LinesNumbered
\SetKwInOut{Input}{Input}
\SetKwInOut{Output}{Output}
\Input{secret message $m$}
\Output{prover key $sk_\lambda$, verifier key $pk_\lambda$}
The certificate authority selects a random $\alpha \in \mathbb{Z}_p$\ for message $m$\;
The certificate authority saves the prover key as $sk_\lambda = \alpha$\;
The certificate authority computes the verifier key as $pk_\lambda = g^{sk_\lambda} \in \mathbb{G}$\;
The certificate authority returns $sk_\lambda$ and $pk_\lambda$ for $\lambda = 1,..., n$.
\caption{\textit{KeyGen}}
\end{algorithm}

\textit{$ProofGen(m, sk_\lambda) \longrightarrow (\Omega_\lambda)$}: This algorithm first computes its hashed identity information $m$, in SHA256 algorithm \cite{rachmawati2018comparative}, as $h = H(m)$. Then, it computes and returns the one-time individual zero-knowledge proofs $\Omega_\lambda = {h}^{sk_\lambda} \in \mathbb{G}$ for $i = 1,..., n$.

\begin{algorithm}[h]
\label{alg: zkp-proofgen}
\SetAlgoLined
\LinesNumbered
\SetKwInOut{Input}{Input}
\SetKwInOut{Output}{Output}
\Input{secret message $m$, prover key $sk_\lambda$}
\Output{one-time ZKP $\Omega_\lambda$ for $\lambda = 1,..., n$}
The truck computes a one-time hash digest $h$ of the secret message $m$
, as $h = H(m)$\;
The truck generates the one-time zero-knowledge proof $\Omega_\lambda = {h}^{sk_\lambda} \in \mathbb{G}$ based on $sk_\lambda$ for $\lambda = 1,..., n$\;
The truck returns $\Omega_\lambda$ for $\lambda = 1,..., n$.
\caption{\textit{ProofGen}}
\end{algorithm}

\textit{$ProofAggregate(\Omega_1, \Omega_2,..., \Omega_n) \longrightarrow (\Omega)$}: This algorithm takes all the individual zero-knowledge proofs as input, computes and returns the aggregated proof $\Omega \longleftarrow \Omega_1\Omega_2\cdot\cdot\cdot\Omega_n$.

\begin{algorithm}[h]
\label{alg: zkp-proofaggregate}
\SetAlgoLined
\LinesNumbered
\SetKwInOut{Input}{Input}
\SetKwInOut{Output}{Output}
\Input{one-time ZKPs $\Omega_\lambda$ for $\lambda = 1,..., n$}
\Output{aggregated one-time ZKP $\Omega$}
The truck computes the aggregated one-time ZKP $\Omega$ as $\Omega = \Omega_1\Omega_2\cdot\cdot\cdot\Omega_n$\;
The truck returns $\Omega$.
\caption{\textit{ProofAggregate}}
\end{algorithm}

\textit{$VerifierKeyAggregate(pk_1, pk_2,..., pk_n) \longrightarrow (pk)$}: This algorithm takes all the verifier keys as input, computes and returns the aggregated verifier key $pk \longleftarrow pk_1pk_2\cdot\cdot\cdot pk_n$.

\begin{algorithm}[h]
\label{alg: zkp-keyaggregate}
\SetAlgoLined
\LinesNumbered
\SetKwInOut{Input}{Input}
\SetKwInOut{Output}{Output}
\Input{verifier keys $pk_\lambda$ for $\lambda = 1,..., n$}
\Output{aggregated verifier key $pk$}
The blockchain computes the aggregated verifier key $pk$ as $pk = pk_1pk_2\cdot\cdot\cdot pk_n$\;
The blockchain returns $pk$.
\caption{\textit{VerifierKeyAggregate}}
\end{algorithm}

\textit{$Verify(\Omega, g, H(m), pk) \longrightarrow (true/false) $}: This algorithm takes the aggregated zero-knowledge proof $\Omega$, the aggregated verifier key $pk$, the generator $g$ and the hashed identity information $H(m)$ as input and check if $e(\Omega, g) = e(H(m), pk)$. Finally, the algorithm returns a Boolean value, either true or false, to validate the aggregated zero-knowledge proof without compromising the identity information of the prover.

\begin{algorithm}[h]
\label{alg: zkp-proofverify}
\SetAlgoLined
\LinesNumbered
\SetKwInOut{Input}{Input}
\SetKwInOut{Output}{Output}
\Input{Aggregated one-time ZKP $\Omega$, aggregated~verifier~key~$pk$, generation $g$, one-time hashed~secret~message $h$}
\Output{verification result $r$}
The blockchain checks
\eIf{$e(\Omega, g) == e(h, pk)$}{$r = true$\;}{$r = false$\;}
The blockchain returns $r$.
\caption{\textit{Verify}}
\end{algorithm}

\subsection{Blockchain Network with Access Control Policy}
In our design, the blockchain functions as the system verifier and provides a distributed ledger that stores the verifier keys and truck platooning records. Data is maintained on-chain to guarantee immutability and integrity. Moreover, the blockchain improves system resilience by providing fault tolerance to the decentralized verification process. Due to fault tolerance and immutability properties, the blockchain serves as a tamper-resistant database recording the system state updates (e.g., proof verification and platoon records). Consequently, the system can tolerate some system outages and lost messages due to faulty or offline blockchain nodes. Furthermore, by storing the platoon history records on-chain, we provide practical benefits to the truck companies. For example, a logistics company can retrieve and analyze the platoon records for their vehicles to determine the optimal platoon size on each route based on historical data. Furthermore, platooning provides additional efficiency and safety benefits, and the platoon records stored on the blockchain can be leveraged to help a company quantify the benefits. 

To protect the sensitive information stored in a given truck's platoon records, we provide programmable access control policies that define what entities can access the on-chain data, and the policies are enforced by our blockchain network. The proposed access control policies are based on the Hyperledger Fabric's access control lists (ACLs)\footnote{https://hyperledger-fabric.readthedocs.io/en/latest/accesscontrol.html}, which can manage access to resources by associating a policy. ACL is similar to the XACML~\cite{nist} that defines a fine-grained, attribute-based access control policy language. By default, a truck company can only retrieve their vehicles' platoon records. This prevents the possibility of a truck profiling attack, where an adversary attempts to reconstruct a truck's driving route by extracting the location data in their platoon records. The specific components involved in our access control scheme are as follows:

\begin{itemize}
\item Participant: It defines the entities involved in the access control procedure. 

\item Operation: It defines the actions governed by the access control policy. Data on our blockchain network is inherently immutable; the supported operations are READ and WRITE. 

\item Resource: It indicates the ledger data to which the access control policy applies to. In our system, the on-chain resources are verifier keys and platoon records.

\item Condition: It defines the conditional statements over multiple variables. Our system can support combinations of multiple conditional logic statements, making it possible to design complex access control policies. 

\item Action: It represents the final decision after executing the access control policy. It can be either ALLOW or DENY.
\end{itemize}

Through the proposed access control scheme, truck companies have complete control over their platoon records, as well as the right to define who can access them. In this way, we provide the desired level of privacy for all participants. This is in stark contrast to traditional centralized systems, where ultimately, the system administrator has full control over the generated data, providing less security, and users generally do not own the rights to their generated data. Simply put, our access control policies define the restrictions regarding who can perform what actions within the blockchain network. For example, the policy defined below represents a rule stating that only the company can READ their trucks' data records:

\begin{verbatim}
rule CompanyCanReadPlatoonRecord {
  description: "Allow company A to read 
        platoon records."
  participant(p): "Company_A"
  operation: READ
  resource(r): "Platoon_Record"
  condition: "r.owner.getIdentifier() === 
        p.getIdentifier()"
  action: ALLOW
}
\end{verbatim}

\section{Correctness and Security Analysis}
Security has paramount importance in the truck platooning systems since vehicular information needs to be handled with proper privacy and also secured against possible attacking vectors. 

\subsection{Correctness Proof}

\noindent
\textbf{Proposition 1.} \textit{The proposed aggregated zero-knowledge proof can correctly verify a truck's identity without revealing the identity information $m$.}
\vspace{1mm}

\begin{proof}
Assume an autonomous truck has the identity information $m$ (e.g., MAC address). It first generates $n$ zero-knowledge proofs $\Omega_\lambda$ for $i = 1,..., n$, and then aggregates these proofs into one short proof as $\Omega$. In this case, $n$ indicates the number of truck companies within the platoon. We will prove that the aggregated proof $\Omega$ can be validated by the $Verify$ algorithm. 

\vspace{0.2cm}
First, multiple individual proofs are generated as:
\begin{align}
   \{H(m), sk_1\} &\longrightarrow \Omega_1 = h^{sk_1}, \\
   \{H(m), sk_2\} &\longrightarrow \Omega_2 = h^{sk_2}, \\
   &\;\;\vdots \notag \\
   \{H(m), sk_n\} &\longrightarrow \Omega_n = h^{sk_n},
\end{align}

Then, the aggregated proof $\Omega$ is computed as:
\begin{equation}
\begin{split}
    \{\Omega_1, \Omega_2,..., \Omega_n\} \longrightarrow \Omega &= \Omega_1\Omega_2\cdot\cdot\cdot\Omega_n \\
    &= h^{sk_1}h^{sk_2}\cdot\cdot\cdot h^{sk_n}, 
\end{split}
\end{equation}

And the aggregated verifier key $pk$ is computed as:
\begin{equation}
\begin{split}
    \{pk_1, pk_2,..., pk_n\} \longrightarrow pk &= pk_1pk_2\cdot\cdot\cdot pk_n \\
    &= g^{sk_1}g^{sk_2}\cdot\cdot\cdot g^{sk_n},
\end{split}
\end{equation}

Next, verifying the aggregated proof $\Omega$ is done by checking that if and only if:
\begin{equation}
\label{equation-verify}
   e(\Omega, g) = e(H(m), pk),
\end{equation}

Now, we prove the Equation (\ref{equation-verify}) based on the bilinear pairing property \cite{eprint-2002-11640}: 
\begin{equation}
\begin{split}
    e(\Omega, g) &= e(\Omega_1\Omega_2\cdot\cdot\cdot\Omega_n, g) \\
    &=e(\Omega_1, g)e(\Omega_2, g)\cdot\cdot\cdot e(\Omega_n, g) \\
    &=e(h^{sk_1}, g)e(h^{sk_2}, g)\cdot\cdot\cdot e(h^{sk_n}, g) \\
    &=e(h, g^{sk_1})e(h, g^{sk_2})\cdot\cdot\cdot e(h, g^{sk_n}) \\
    &=e(h, pk_1)e(h, pk_2)\cdot\cdot\cdot e(h, pk_n) \\
    &=e(h, pk_1pk_2\cdot\cdot\cdot pk_n) \\
    &=e(h, pk)\\
    &=e(H(m), pk).
\end{split}    
\end{equation}

\noindent
\textbf{Bilinear Pairing Property:} \textit{Let $\mathbb{G}$ be a multiplicative cyclic group of prime order $p$ with generator $g$. Let $e : \mathbb{G} \times \mathbb{G} \rightarrow \mathbb{G}_T$ be a computable, bilinear and non-degenerate pairing into the group $\mathbb{G}_T$. Then, we have $e(x^a, y^b) = e(x, y)^{ab}$ for all $x,y \in \mathbb{G}$ and $a,b \in \mathbb{Z}_p$ because $\mathbb{G}$ is cyclic.}

\end{proof}

\subsection{Security Analysis}
The maximal privacy-preserving is one of the most significant security requirements of the proposed authentication system for truck platooning, which indicates that  attackers should not be capable of revealing the sensitive message in the aggregated zero-knowledge proof.

\subsubsection{Threat Model}
The threat model for maximal privacy-preserving is formally defined by the interaction game between adversary $A$ and challenger $C$. Let the number of truck companies be $n$. In a mixed fleet vehicular network, the compromise of all $n$ truck companies will directly lead to the revealing of any autonomous truck based on their registered information. Thus, we assume that the adversary $A$ in the security game is able to control at most $n-1$ truck companies. The adversary $A$ can adaptively query the key pairs between autonomous trucks and corrupted truck companies, and the aggregated zero-knowledge proofs. A truck $\chi$ from the uncorrupted company generates the aggregated zero-knowledge proof $\Omega$ for its secret message $m$ (e.g., MAC address). Then, $A$ sends $\Omega$ to $C$ as a challenge. As a response, $C$ generates the aggregated verifier key and makes the authentication for $\Omega$. If the authentication passes, $A$ should reveal the secret message $m$ of the truck $\chi$.

\textbf{Setup.} $C$ creates prover/verifier key pairs for the truck $\chi$, which comes from the only one uncorrupted company in the mixed fleet vehicular network.

\textbf{ZKP Generation.} The truck $\chi$ generates $n$ one-time zero-knowledge proofs $\Omega_\lambda$ ($\lambda = 1, 2, \cdots, n$) and aggregates them as one proof $\Omega$.

\textbf{Queries.} $A$ adaptively makes the following queries.

\textit{$-$Key pair queries.} $A$ queries on the key pairs between the truck $\chi$ and corrupted truck companies.

\textit{$-$Aggregated ZKP query.} $A$ queries for the aggregated zero-knowledge proof $\Omega$ for the truck $\chi$.

\textbf{Challenge.} $A$ sends $\Omega$ to $C$ as a challenge. The request is that the key pair between the truck $\chi$ and the uncorrupted company has not been queried. Then, $C$ generates the aggregated verify key $pk$ and makes the authentication for the truck $\chi$. If the authentication passes, $C$ notifies $A$ to make the following guess.

\textbf{Guess.} $A$ outputs a guess $m'$ to reveal the secret message of truck $\chi$. If $m' = m$, the challenger $C$ outputs $1$ meaning the adversary $A$ wins the game. Otherwise, $C$ outputs $0$.

\subsubsection{Security Claim}
\noindent
\textbf{Proposition 2. } \textit{If the zero-knowledge of the proposed aggregated zero-knowledge proof holds, then the system achieves maximal privacy-preserving for truck platooning in a mixed fleet vehicular network.}
\vspace{1mm}

\begin{proof}
As defined in the threat model, we assume that the adversary $A$ has the capability to control $n-1$ truck companies, queries the key pairs between the truck $\chi$ and the corrupted companies, and even intercepts the aggregated zero-knowledge proof $\Omega$ of the truck $\chi$ (e.g., eavesdropping attack). However, the adversary $A$ can not reuse the proof $\Omega$ to spoof the verification process (e.g., fake identity attack and replay attack) or reveal the original sensitive message $m$ from $\Omega$. 

First, as shown in Equation 14, verifying the aggregated proof $\Omega$ is done by checking if and only if $e(\Omega, g) = e(H(m), pk)$, in which the original message $m$ is hidden during the verification process. Thus, the adversary $A$ can not reveal the original message $m$ from $\Omega$. Second, the verification algorithm needs a correct pair of $\Omega$ and $H(m)$, which both are one-time generated for use. In addition, without knowing the key pair between the truck $\chi$ and the uncorrupted company, the adversary $A$ can not derive the current hash digest $H(m)$ based on Equation 12. As a result, even if the adversary $A$ intercepts the proof $\Omega$, $A$ can not reuse it for the next round of verification because $H(m)$ will be updated.
\end{proof}

\section{Implementation and Evaluation}

\subsection{Implementation}

We implement the proposed authentication system and conduct a series of experiments to evaluate its performance. The system consists of two primary portions that interact seamlessly: the verification module based on aggregated zero-knowledge proof and the blockchain network. The aggregated ZKP scheme is programmed by using the Hyperledger Ursa library\footnote{https://www.hyperledger.org/projects/ursa}. The blockchain network is developed on the Hyperledger Fabric platform\footnote{https://www.hyperledger.org/projects/fabric} and tested using the Hyperledger Caliper benchmark tool\footnote{https://www.hyperledger.org/use/caliper}. We then instantiate multiple participants in the blockchain network, including different autonomous trucks and companies, and perform a variety of experiments having varying endorsement policies and transaction send rates. The endorsement policy signifies the number of peers which must verify the legitimacy of a transaction, while the transaction send rates indicates the number of new transactions per second. The prototype and experiments are deployed and conducted on multiple Fabric peers in Docker containers on Ubuntu 18.04 operating system with 2.8 GHz Intel i5-8400 processor and 8GB DDR4 memory.

\subsubsection{Aggregated Zero-Knowledge Proof}
\label{ursa-experiment}

\begin{figure}[t]
\centering
\includegraphics[width=0.45\textwidth]{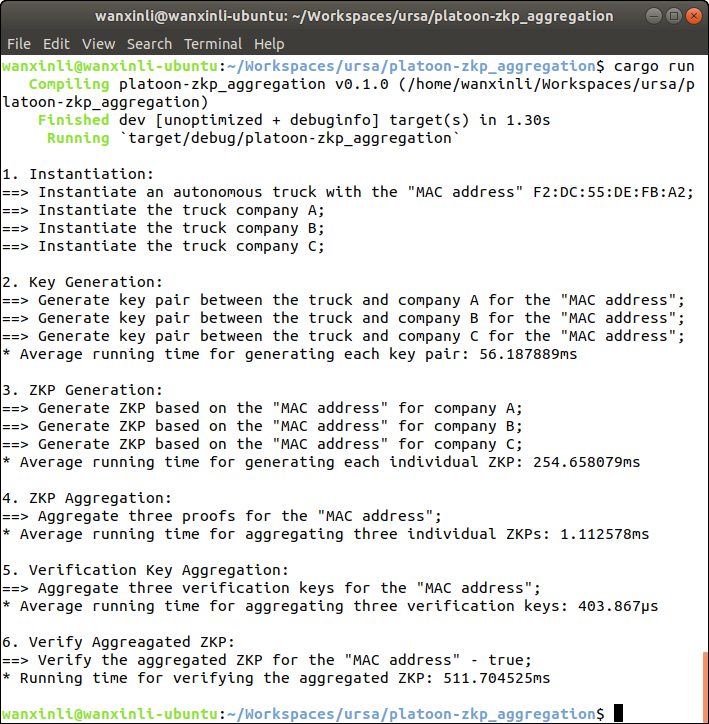}
\caption{Process of the Aggregated ZKP on Hyperledger Ursa.}
\label{fig:ursa-process}
\end{figure}

As illustrated in Fig. \ref{fig:ursa-process}, the aggregated ZKP scheme performs multiple functions including key generation, zero-knowledge proof generation, aggregations of zero-knowledge proofs and verifier keys, and verification of the aggregated ZKP. These functionalities are programmed by using Hyperledger Ursa, a cryptographic library for Hyperledger applications. Hyperledger Ursa is programmed using the Rust language and provides APIs for various cryptographic schemes. Our verification module operates in the following six phases:

\vspace{0.2cm}
Phase 1 - Initialization: Phase 1 initializes the participant instances including an autonomous truck and three truck companies. In this example, the autonomous truck has the identifier information {\tt mac\_address} (value: {\tt F2:DC:55:DE:FB:A2}).

\vspace{0.2cm}
Phase 2 - Key Generation: The certificate authority generates the key pairs between this autonomous truck and each company in this phase. The average running time for generating each key pair takes 56 ms.

\vspace{0.2cm}
Phase 3 - ZKP Generation: In Phase 3, the autonomous truck applies the prover keys to generate three individual zero-knowledge proofs, one for each of the three companies. The average running time for this signing process is about 255 ms.

\vspace{0.2cm}
Phase 4 - ZKP Aggregation: In this phase, the three individual proofs generated from Phase 3 are aggregated into one short proof. Our results indicate that the average running time for aggregating three signatures is about 1.1 s.

\vspace{0.2cm}
Phase 5 -- Verification Key Aggregation: In Phase 5, the module automatically aggregates multiple verification keys from Phase 2 into one short verifier key. The result indicates that the average running time for aggregating three verifier keys is around 0.4 ms.

\vspace{0.2cm}
Phase 6 -- Verify Aggregated ZKP: Finally, the blockchain peers can verify the aggregated ZKP using the aggregated verifier key from Phase 5. The verification algorithm takes the aggregated proof, the aggregated verification key, the associated generator and the hashed identifier value as inputs and uses bilinear pairing to verify aggregated ZKP. The average running time for verifying one aggregated ZKP is about 512 ms.

\subsubsection{Blockchain Network with Access Control Policy}
\label{blockchain-network}
Hyperledger Fabric is an open-source and modular permissioned blockchain framework. The four programmable modules used in our system are: model file (.cto) which is used to define all of the data structures in the network; script file (.js) where smart contracts are written; access control list (.acl) for deploying access control policies; and the query file (.qry) which defines the query operations similarly to a traditional database system.

In our prototyped blockchain system, we provide a web portal for the network participants (autonomous trucks and companies), which can be used to interact with the blockchain network. An example is shown in Fig. \ref{fig:login-window}, where each participant has a registered ID for connecting to the blockchain. The trusted entity operating as the certificate authority in our system (e.g., DMV) also acts as the blockchain administrator, issuing access control permissions for autonomous trucks and companies.

\begin{figure}[b]
\centering
\includegraphics[width=0.45\textwidth]{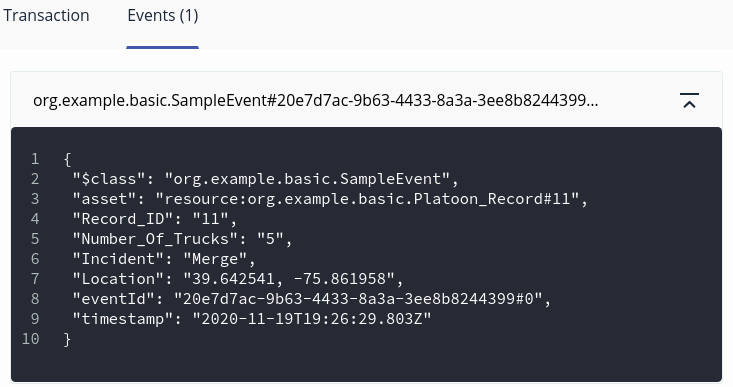}
\caption{A company can retrieve its owned truck's platoon record from the blockchain ledger.}
\label{fig:truck-info}
\end{figure}

We construct our blockchain prototype and perform multiple experiments on the access control policies. Once the defined access control policy is executed and verified, the client will be able to retrieve the on-chain data in the form of historical transactions. An example of platoon record retrieval is shown in Fig. \ref{fig:truck-info}. After any successful retrieval, the blockchain network stores the event in an access log with the timestamp. Our access control policies prevent client users from accessing others' transactions, unlike in a permissionless blockchain system where all users can access all historical transactions. For example, if an attacker, or another user not included in the defined access control policy, attempts to retrieve a record from another user, our blockchain network will reject this request immediately.

\textbf{\begin{figure}[t]
\centering
\includegraphics[width=0.45\textwidth]{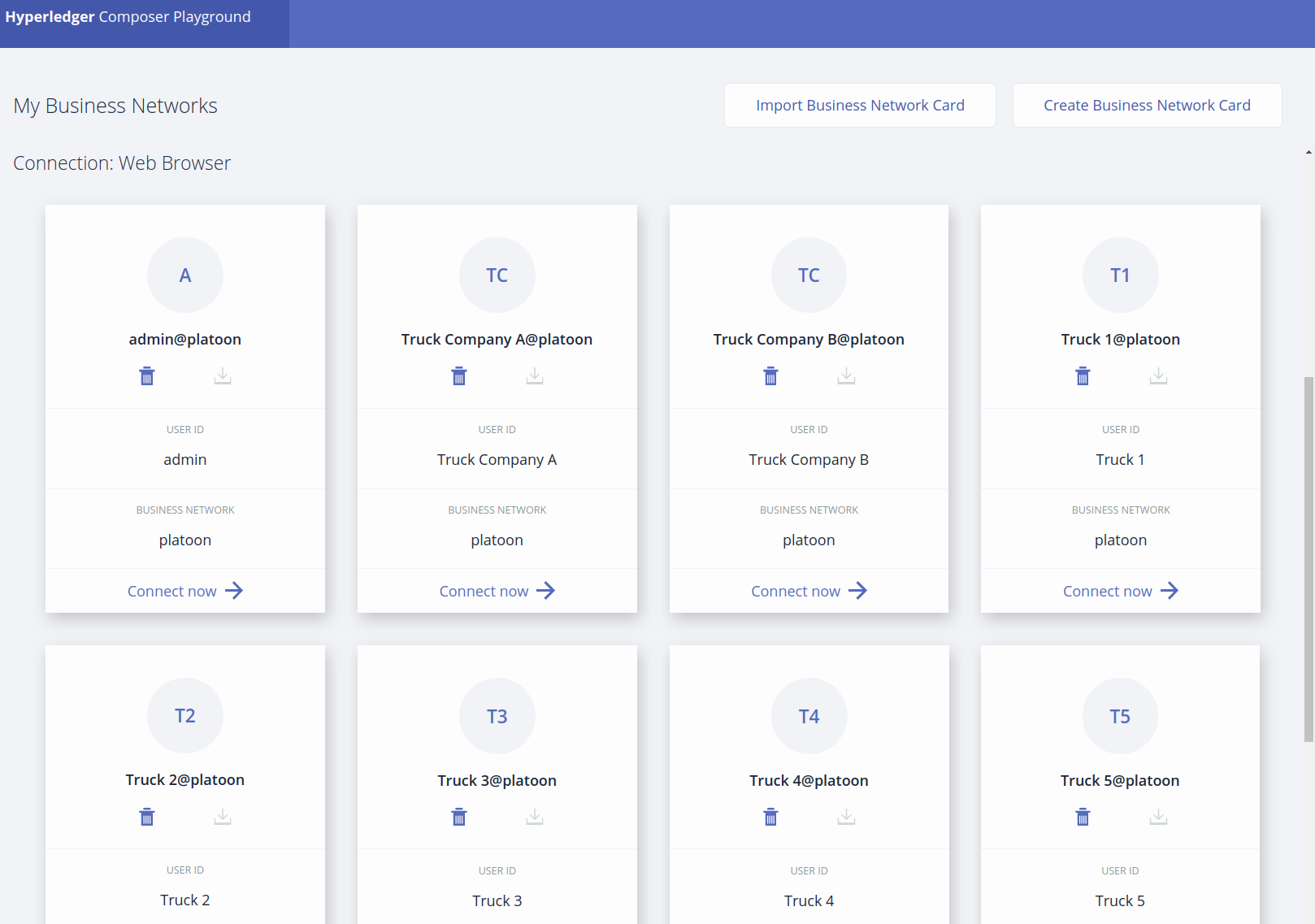}
\caption{Blockchain network login window for companies and autonomous trucks.}
\label{fig:login-window}
\end{figure}}

\subsection{Experimental Results}

\subsubsection{Aggregated ZKP Running Time}
\begin{figure}[t]
\centering
\includegraphics[width=0.35\textwidth]{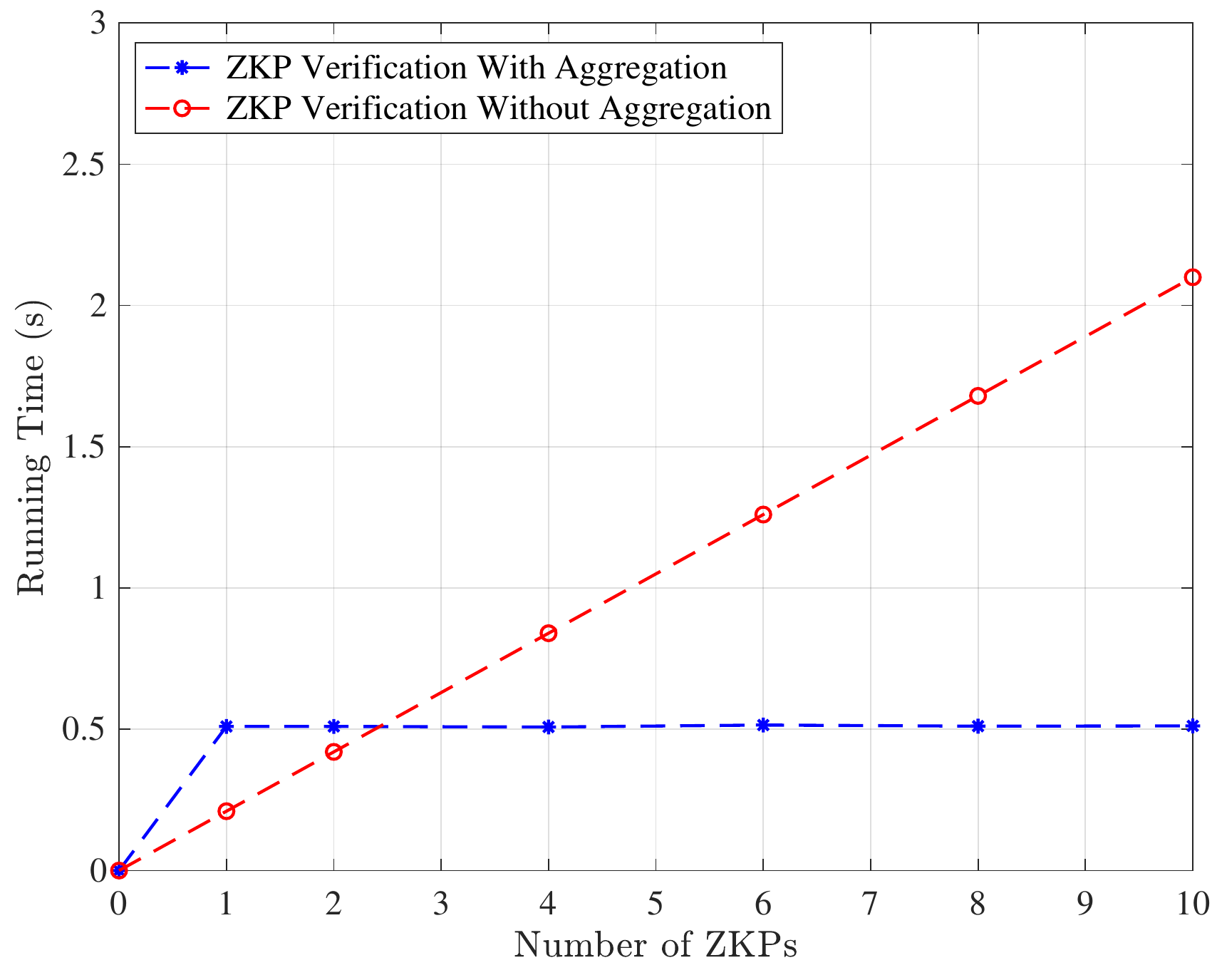}
\caption{Running time comparison of ZKP verification with aggregation and ZKP verification without aggregation when increasing the number of ZKPs from 0 to 10.}
\label{fig:ursa_verification}
\end{figure}

In order to quantify the performance of our aggregated ZKP scheme, we conduct multiple experiments and measure the running time for both the proof verification and aggregation stages. Fig. \ref{fig:ursa_verification} compares the running time for verification of an aggregated proof with that of our previously non-aggregated ZKP scheme \cite{li2021icbc}, where the x-axis represents the number of proofs to be verified. In our previous scheme, proofs are verified sequentially. The results show that the running time for the non-aggregated ZKP scheme scales linearly with the number of proofs to be verified, while our aggregated approach can offer a constant verification time of 500 milliseconds regardless of the number of proofs. This is because the verification latency of an aggregated ZKP becomes independent of the number of proofs, offering a significant performance improvement when the proof count exceeds two.

\begin{figure}[t]
\centering
\includegraphics[width=0.35\textwidth]{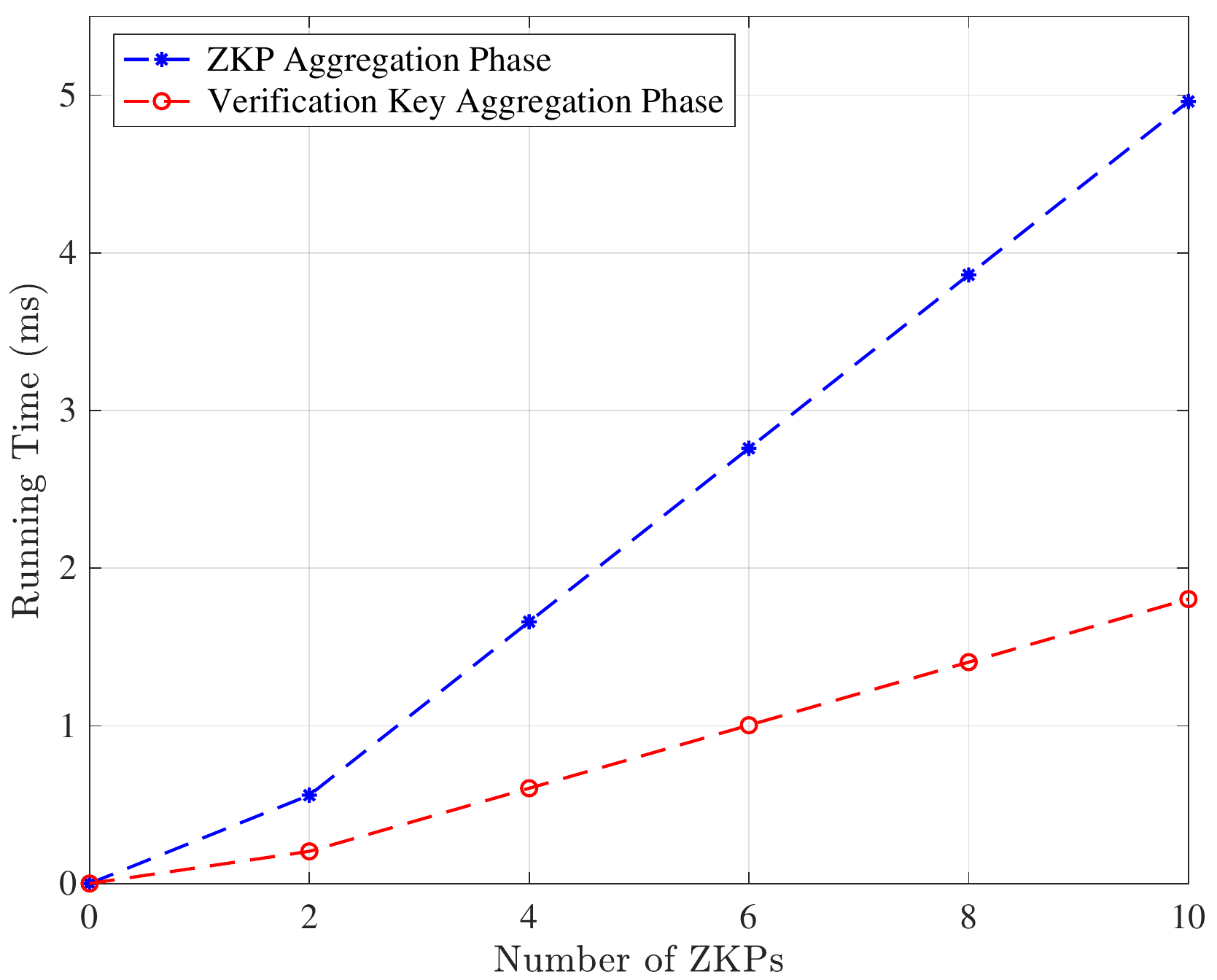}
\caption{Running time comparison of ZKP aggregation phase and verification key aggregation phase when increasing the number of ZKPs from 0 to 10.}
\label{fig:ursa_aggregation}
\end{figure}

Additionally, Fig. \ref{fig:ursa_aggregation} illustrates the running time for the aggregation-related functions, including the initial ZKP aggregation stage (by the prover) and the verification key aggregation stage (by the verifier), with respect to the number of ZKPs to be verified. The results show that our new aggregation scheme can perform both stages in a few milliseconds, providing a desirable response time for real-world truck platooning. 
Moreover, when considering the combined running time of the aggregation-related functions with that of the ZKP verification phase, 
the time for aggregation-related functions becomes negligible compared to the overall running time.

\subsubsection{Transaction Throughput}
\begin{figure}[t]
\centering
\includegraphics[width=0.35\textwidth]{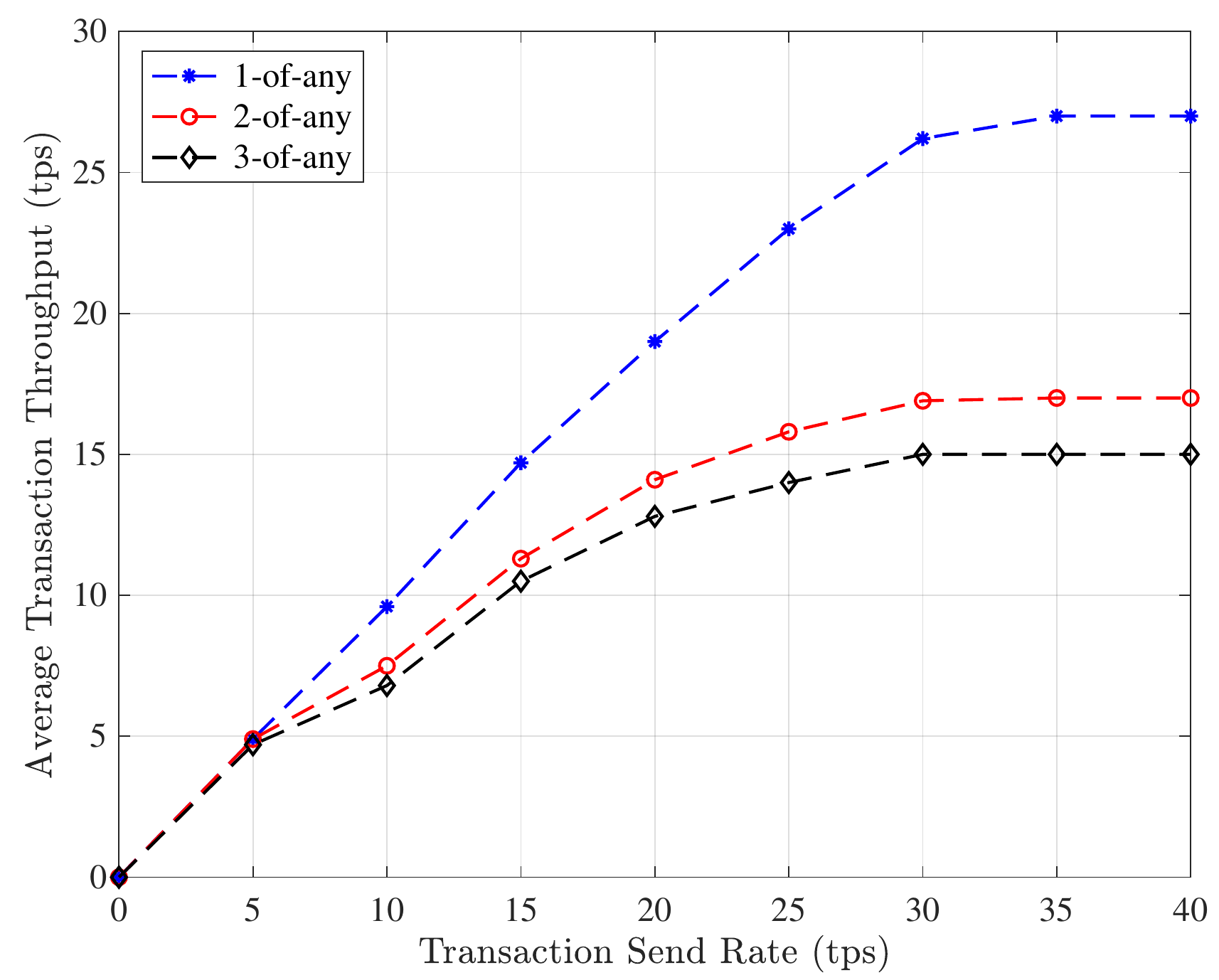}
\caption{Transaction throughput comparison of different Hyperledger Fabric endorsement policies when increasing the transaction send rates from 0 to 40 tps.}
\label{fig:throughput_sendrate}
\end{figure}

\begin{figure}[t]
\centering
\includegraphics[width=0.35\textwidth]{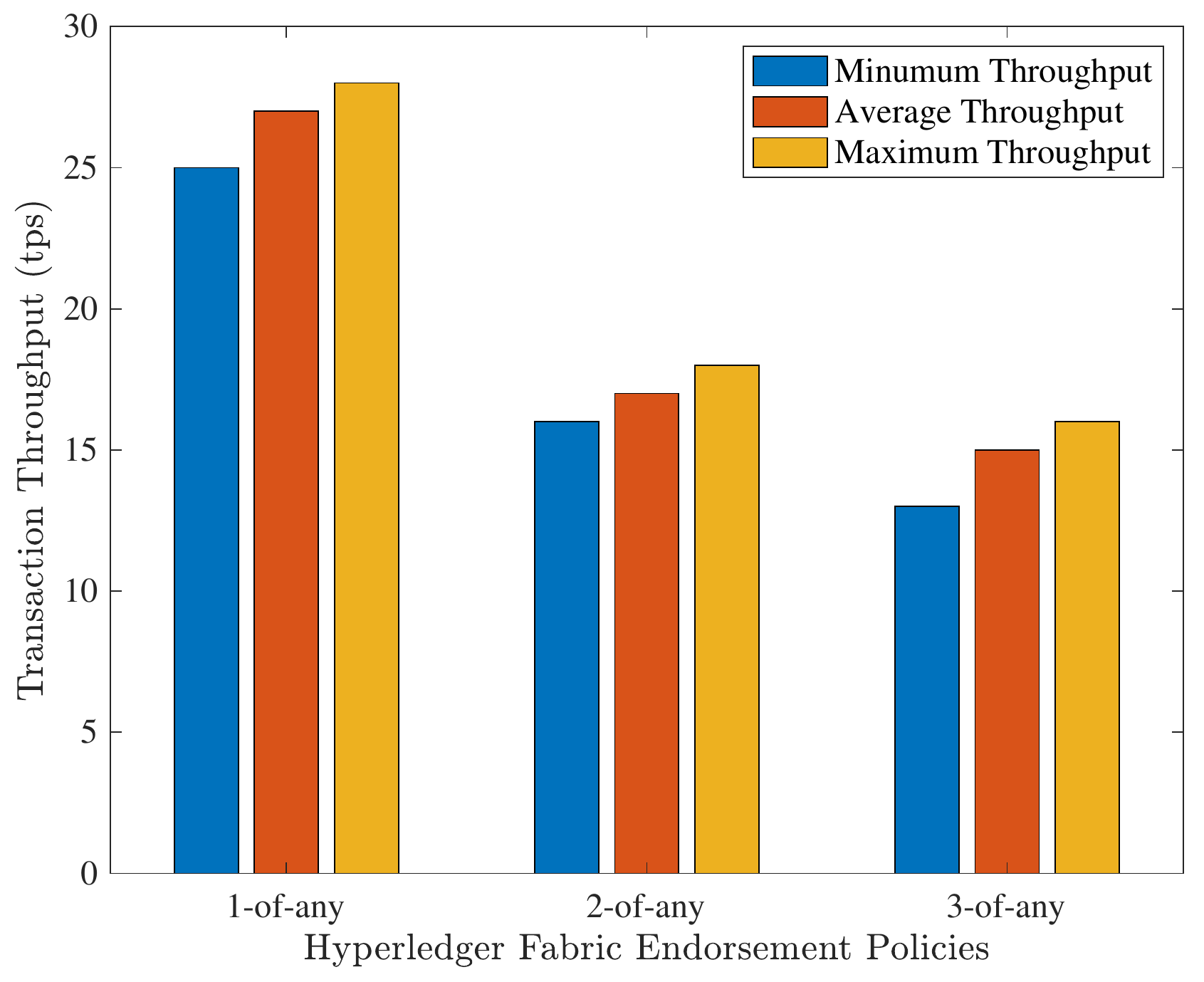}
\caption{Minimum, average and maximum transaction throughputs of different Hyperledger Fabric endorsement policies under the transaction send rate of 30 tps.}
\label{fig:tps_endorse}
\end{figure}

Transaction throughput for a blockchain network quantifies the rate at which transactions are processed through the network over a given time cycle, in units of transactions per second. As shown in Fig. \ref{fig:throughput_sendrate} and Fig. \ref{fig:tps_endorse}, we evaluate the throughput results under different transaction send rates and Hyperledger Fabric endorsement policies. In Fig. \ref{fig:throughput_sendrate}, the average transaction throughput will increase at the beginning and peaks at 27 tps, 17 tps and 15 tps under 1-of-any, 2-of-any and 3-of-any endorsement policies, respectively. When the system reaches its peak performance, the overloaded transactions will be queued.

The endorsement policy impacts on transaction throughput because it signifies the number of peers which must verify the legitimacy of a transaction. In Fig. \ref{fig:tps_endorse}, our results show that the number of endorsing peers has an inverse relationship to the network's transaction throughput. This is because increasing the number of peers required to validate a transaction also increases the complexity of the endorsement process. That being said, our results 
show that the performance is relatively stable for a given endorsement policy, and the difference between the minimum, maximum and average cases is minor.

\subsubsection{Transaction Latency}
\begin{figure}[t]
\centering
\includegraphics[width=0.35\textwidth]{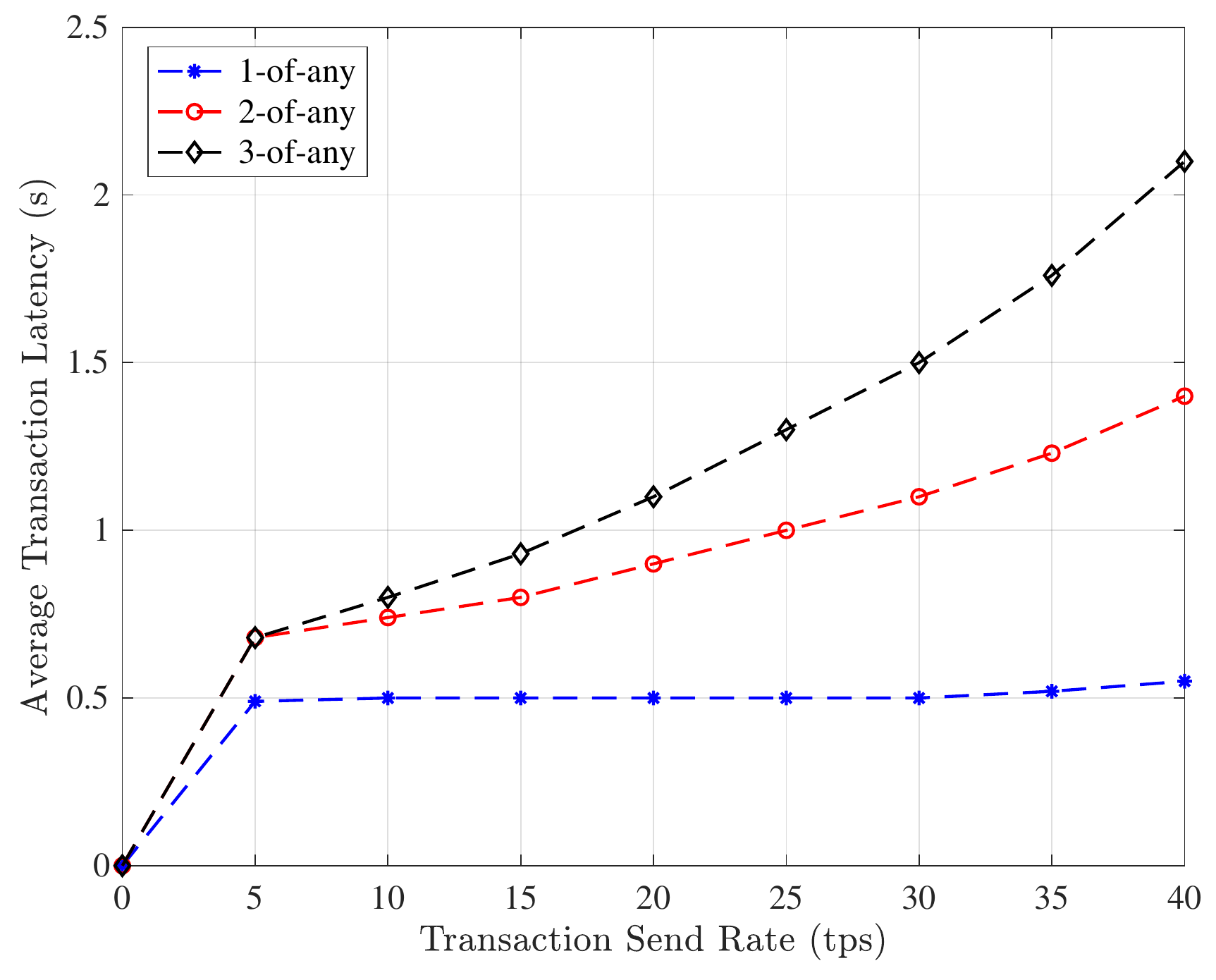}
\caption{Transaction latency comparison of different Hyperledger Fabric endorsement policies when increasing the transaction send rates from 0 to 40 tps.}
\label{fig:latency_sendrate}
\end{figure}

\begin{figure}[t]
\centering
\includegraphics[width=0.35\textwidth]{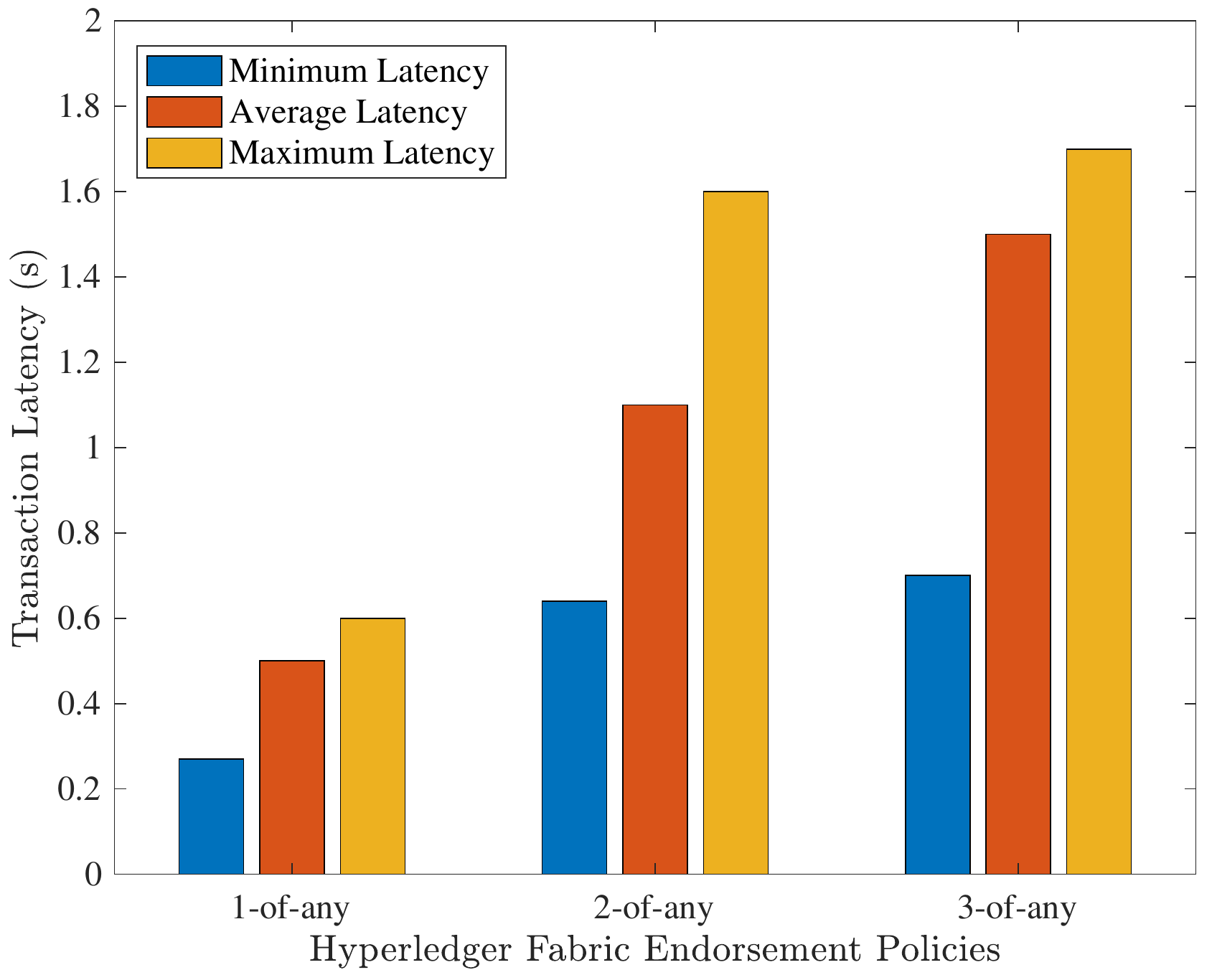}
\caption{Minimum, average and maximum transaction latencies of different Hyperledger Fabric endorsement policies under the transaction send rate of 30 tps.}
\label{fig:latency_endorse}
\end{figure}

We also perform experiments to measure and quantify the transaction latency of our prototyped blockchain network. Transaction latency measures the processing time for a blockchain transaction, from client submission to when the transaction is committed to the ledger. We perform multiple experiment rounds with varying transaction send rates and Hyperledger Fabric endorsement policies and compiled our results in Fig. \ref{fig:latency_sendrate} and Fig. \ref{fig:latency_endorse}. In Fig. \ref{fig:latency_sendrate}, when increasing the transaction send rate, the average transaction latency will increase under 2-of-any and 3-of-any endorsement policies but remain relatively constant around 0.5s under 1-of-any endorsement policy. 

As shown in Fig. \ref{fig:latency_endorse}, the relationship between the transaction latency and endorsement policy is readily apparent: as the number of endorsing peers increases, we see an increase in both average and maximum transaction latency. However, the increase in latency when changing the endorsement policy from 2-of-any to 3-of-any is significantly lower than when moving from a 1-of-any to a 2-of-any policy. 

\subsubsection{Execution Time of Platooning Formation}
We present the performance of platooning formation phase, which contains the \emph{1st catch-up} and \emph{cooperative driving} steps. 
In the experiments, we set the maximum vehicle-to-vehicle communication radius $R = 300 m$, 
initial speed of the standalone truck $v_{\chi}(t) = 60 km/h = 17 m/s$, acceleration rate of the standalone truck $a_{\chi}(t) = 1 m/{s^2}$ and maximum speed of the standalone truck $v_{\chi}^{max} = 100 km/h = 28 m/s$, initial and maximum speed of the platoon $v_{i}(t) = v_{i}^{max} = 80 km/h = 22 m/s$, deceleration rate of the platoon $a_{i}(t) = - 1 m/{s^2}$. To compare the execution time between the \emph{2nd catch-up} and \emph{slow-down} strategies, we set $|a_{\chi}(t)| = |a_{i}(t)|$ as default. And we set the Hyperledger Fabric endorsement policy as \emph{2-of-any} to obtain the average blockchain network latency from Fig. \ref{fig:latency_endorse}, which is $1.1 s$.

\begin{figure}[t]
\centering
\includegraphics[width=0.35\textwidth]{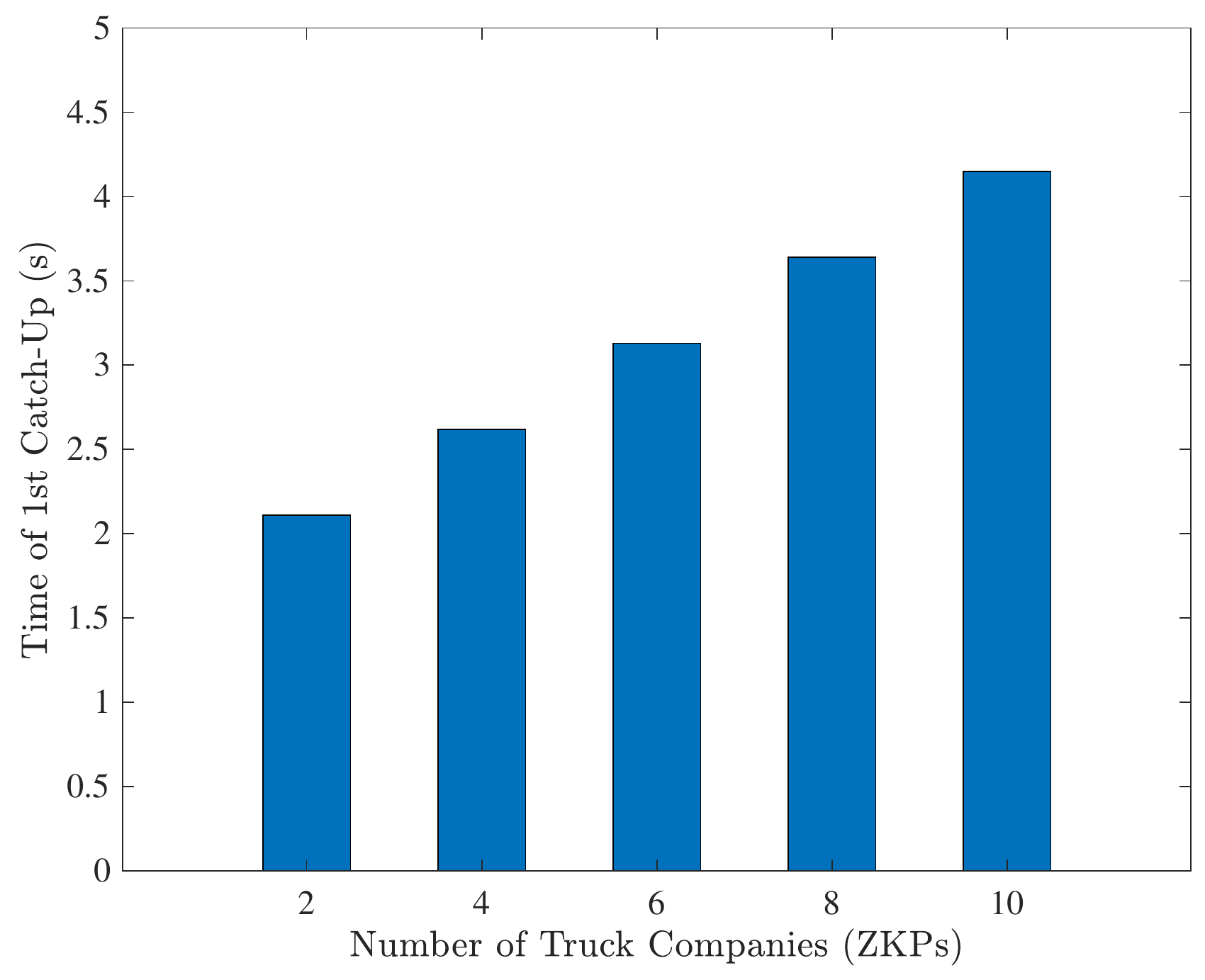}
\caption{Execution time of the 1st catch-up step when increasing the number of truck companies from 2 to 10 in the vehicular network.}
\label{fig:1st catch-up}
\end{figure}

We vary the number of truck companies in the network from 2 to 10, which requires the standalone truck to generate one-time zero-knowledge proofs from 2 to 10 before the proof aggregation step (refer to Fig. \ref{fig:workflow}), respectively. First, we record the \emph{1st catch-up} time $\Gamma$ (end-to-end authentication time). As shown in Fig. \ref{fig:1st catch-up}, with increasing the number of zero-knowledge proofs, $\Gamma$ will slightly increase from $2.1s$ to $4.2s$ due to more time needed for generating one-time ZKPs. More specifically, generating a one-time ZKP will add a time cost of $255 ms$ to $\Gamma$. 

\begin{figure}[t]
\centering
\includegraphics[width=0.35\textwidth]{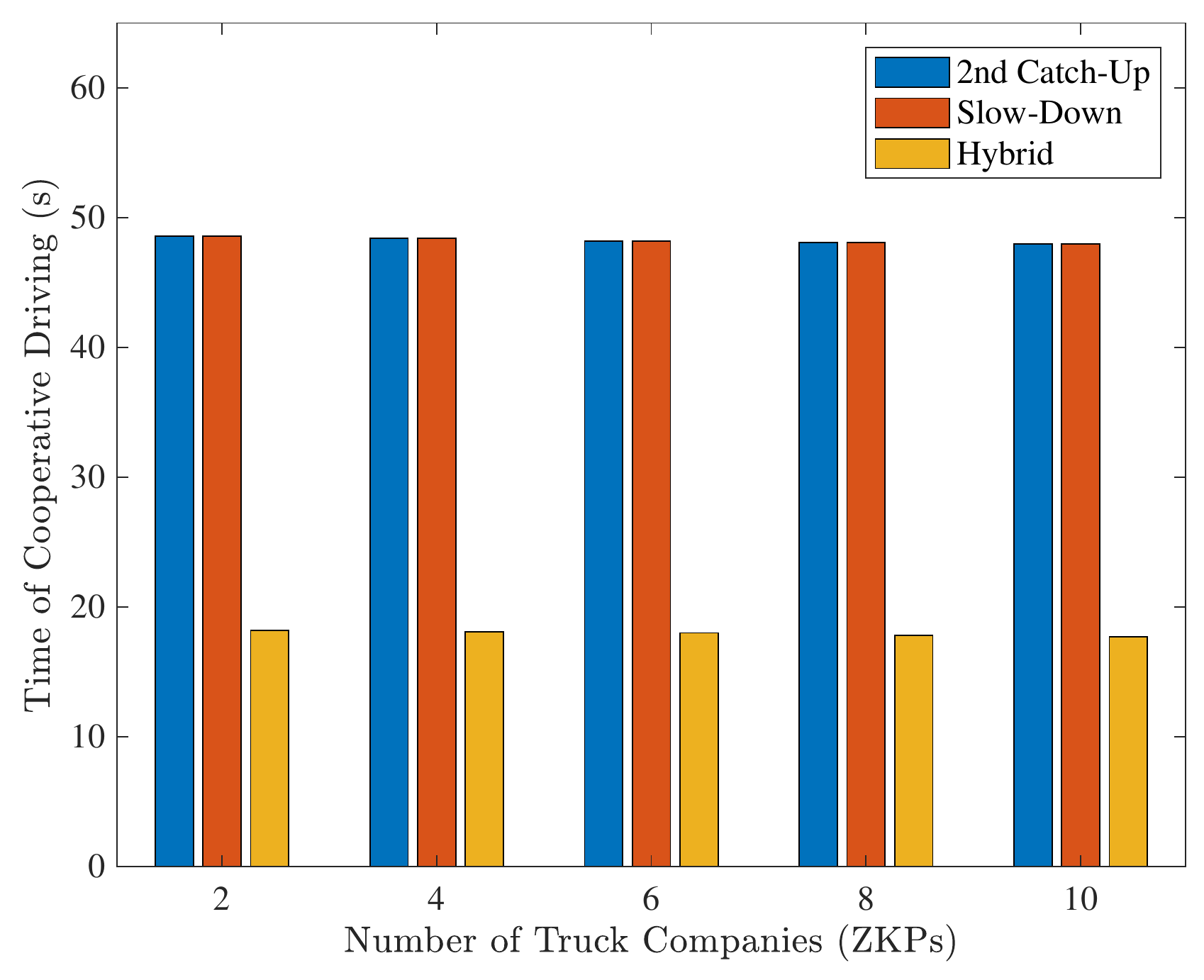}
\caption{Execution time of the cooperative driving step when increasing the number of truck companies from 2 to 10 in the vehicular network under different cooperative driving strategies (2nd catch-up, slow-down and hybrid).}
\label{fig:cooperative driving}
\end{figure}

\begin{figure}[t]
\centering
\includegraphics[width=0.35\textwidth]{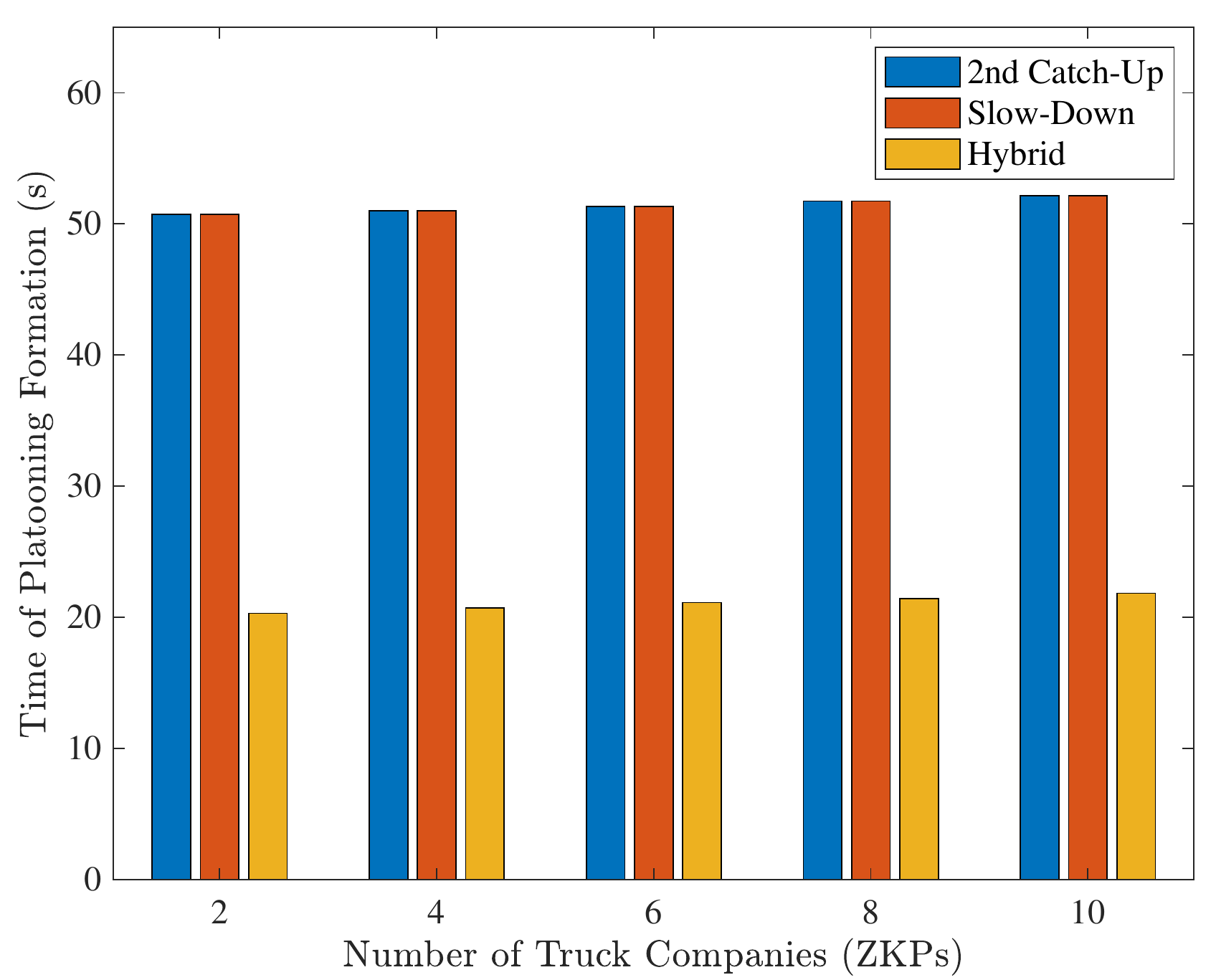}
\caption{Total execution time of the platooning formation when increasing the number of truck companies from 2 to 10 in the vehicular network under different cooperative driving strategies (2nd catch-Up, slow-down and hybrid).}
\label{fig:total}
\end{figure}

Next, we calculate the \emph{cooperative driving} time $\Theta$ based on Equation \ref{eq:platoonStrategy2} and the total platooning formation time $\mathrm{T}$ based on Equation \ref{eq:total}. As shown in Fig. \ref{fig:cooperative driving}, the \emph{2nd catch-up} and \emph{slow-down} strategies have same time cost when the acceleration rate of the standalone truck equals the deceleration rate of the platoon ($|a_{\chi}(t)| = |a_{i}(t)|$). The \emph{hybrid} strategy has better performance in time cost than the \emph{2nd catch-up} and \emph{slow-down} strategies because both the standalone truck and the platoon move towards an intermediate state. In comparison with the \emph{1st catch-up} step, \emph{cooperative driving} will take more time, especially when a long distance of vehicle-to-vehicle communication radius $R$ is given. Besides, $\Theta$ remains constant when more truck companies are considered due to the aggregated ZKP authentication protocol. Fig. \ref{fig:total} presents the total platooning formation time $\mathrm{T}$. With increasing the number of zero-knowledge proofs, $\mathrm{T}$ will have a negligible increase by considering both \emph{1st catch-up} and \emph{cooperative driving} steps.

\subsubsection{Resource Consumption}

Throughout our experiments, we collect data on the resource consumption for each node across all three endorsement policies, and the results are shown in Table \ref{tabble:consumption}. As the number of peers in the endorsement policy increases, the network traffic also increases. This is because, by adding more peers to the endorsement policy, we elevate the complexity of the process and require more communication overhead for each endorsement. 

\begin{table}[t]
\caption{Resource Consumption}
\label{tabble:consumption}
\begin{tabular}{|c|c|c|c|c|c|}
\hline
\textbf{Type}                      & \textbf{Peer}  & \textbf{Memory}  & \textbf{CPU}     & \textbf{Traffic In} & \textbf{Traffic Out} \\ \hline
1-of-any                  & peer1 & 202.3MB & 15.11\% & 862.3KB    & 880.4KB     \\ \hline
\multirow{2}{*}{2-of-any} & peer1 & 209.5MB & 12.51\% & 984.1KB    & 903.4KB     \\ \cline{2-6} 
                          & peer2 & 244.7MB & 13.18\% & 967.1KB    & 567.8KB     \\ \hline
\multirow{3}{*}{3-of-any} & peer1 & 200.5MB & 11.48\% & 1.1MB      & 881.0KB     \\ \cline{2-6} 
                          & peer2 & 245.0MB & 11.76\% & 1.1MB      & 547.4KB     \\ \cline{2-6} 
                          & peer3 & 223.3MB & 12.26\% & 1.1MB      & 549.2KB     \\ \hline
\end{tabular}
\end{table}

\section{Conclusion}

This paper introduces an aggregated and efficient zero-knowledge proof approach to privacy-preserving identity verification atop a permissioned blockchain network for authentication in the mixed fleet platooning environment. We provide the correctness proof and the security analysis of our proposed authentication scheme, highlighting its increased security and fast performance in comparison to a single proof design. The blockchain performs the role of verifier within the authentication scheme, reducing unnecessary communication overhead. Moreover, the blockchain improves system resilience by providing fault tolerance to the decentralized verification process. Platooning records are stored directly on the digital ledger to guarantee the data immutability and integrity, while our programmable access control policies ensure data privacy. To evaluate the end-to-end authentication time, we implement our proposed scheme using the Hyperledger platform and conduct extensive benchmark tests. The results demonstrate that our proposed approach can perform authentication on the order of milliseconds, regardless of the number of proofs, highlighting feasibility for real-world deployment.

\section*{Acknowledgment}
This paper is a revised and extended version of \cite{li2021icbc} published at the IEEE International Conference on Blockchain and Cryptocurrency (IEEE ICBC 2021). We thank Zijia Zhong (New Jersey Institute of Technology) and Gene Donaldson (DelDOT TMC) for helpful discussions and insights. This research was partially supported by a Federal Highway Administration grant: ``Artificial Intelligence Enhanced Integrated Transportation Management System", 2020-2023. This research was also partially supported by the Guangdong Basic and Applied Basic Research Foundation under Grant No. 21201910240004714, and Natural Science Foundation of Shaanxi Provincial Department of Education under Grant No. 2022JQ-639, and the Basic Research Programs of Taicang under Grant No. TC2022JC23.

\Urlmuskip=0mu plus 1mu\relax
\bibliographystyle{IEEEtran}
\bibliography{sig.bib}

\section*{Biographies}
\vskip -1\baselineskip plus -1fil 

\begin{IEEEbiography}
[{\includegraphics[width=1in,height=1.25in,clip]{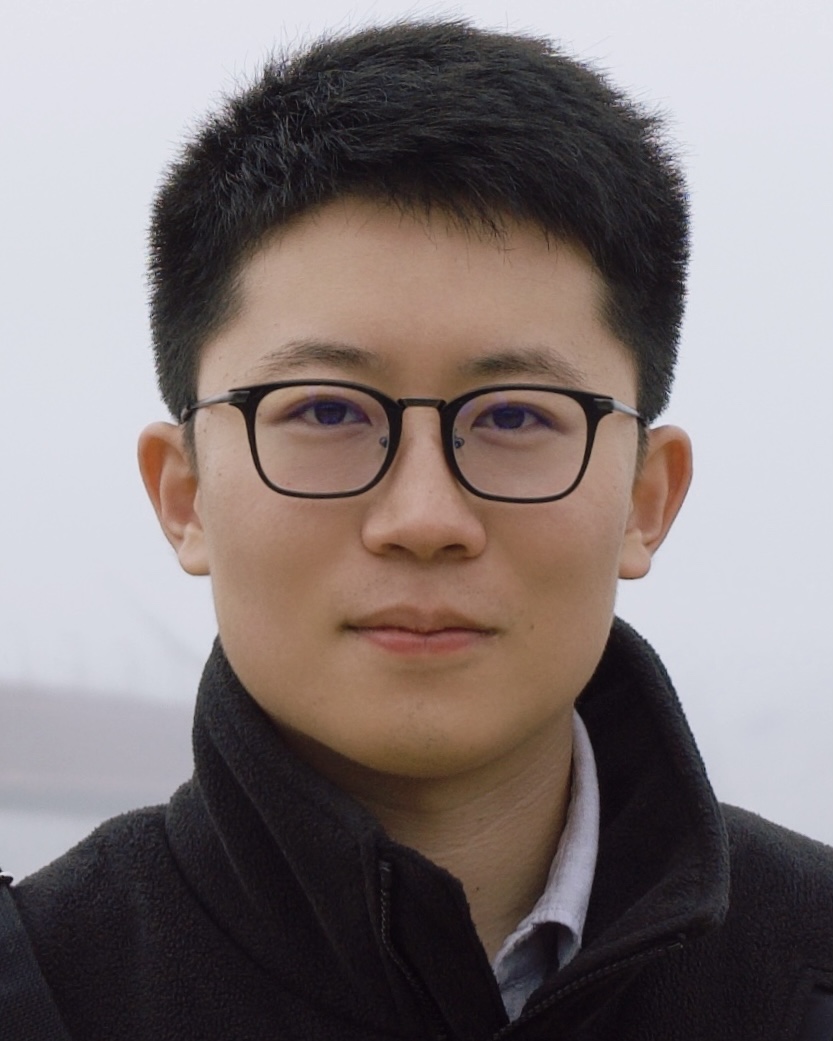}}]{Wanxin Li}(Member, IEEE) received the B.S. degree from Chongqing University in 2015, and the M.S. and Ph.D. degrees from the University of Delaware in 2017 and 2022, respectively. He is currently a Lecturer in the Department of Communications and Networking at Xi'an Jiaotong-Liverpool University. His research interests include blockchain, cryptography, machine learning and smart transportation. He was honored with the 2022 Best Dissertation Award from IEEE ITSS. 
\end{IEEEbiography}
\vskip -6pt plus -1fil

\begin{IEEEbiography}
[{\includegraphics[width=1in,height=1.25in,clip]{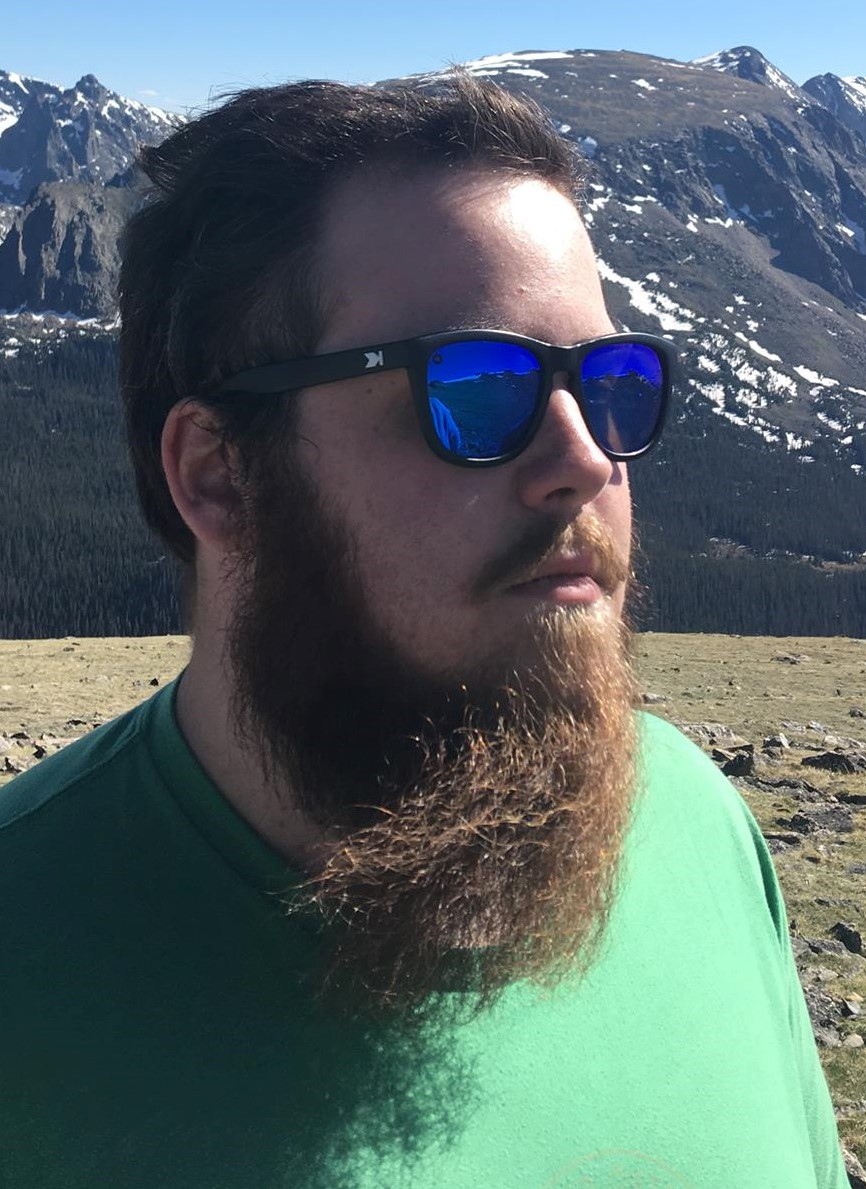}}]{Collin Meese} (Member, IEEE) received the B.S. degree in computer science from the University of Delaware in 2020. He is currently working toward a Ph.D. degree at the University of Delaware. He received the NSF Graduate Research Fellowship award in 2022. His research interests include blockchain, vehicular networks, distributed machine learning, connected and autonomous vehicles, and intelligent civil systems. 
\end{IEEEbiography}

\vskip -6pt plus -1fil
\begin{IEEEbiography}
[{\includegraphics[width=1.0in,height=1.25in,clip]{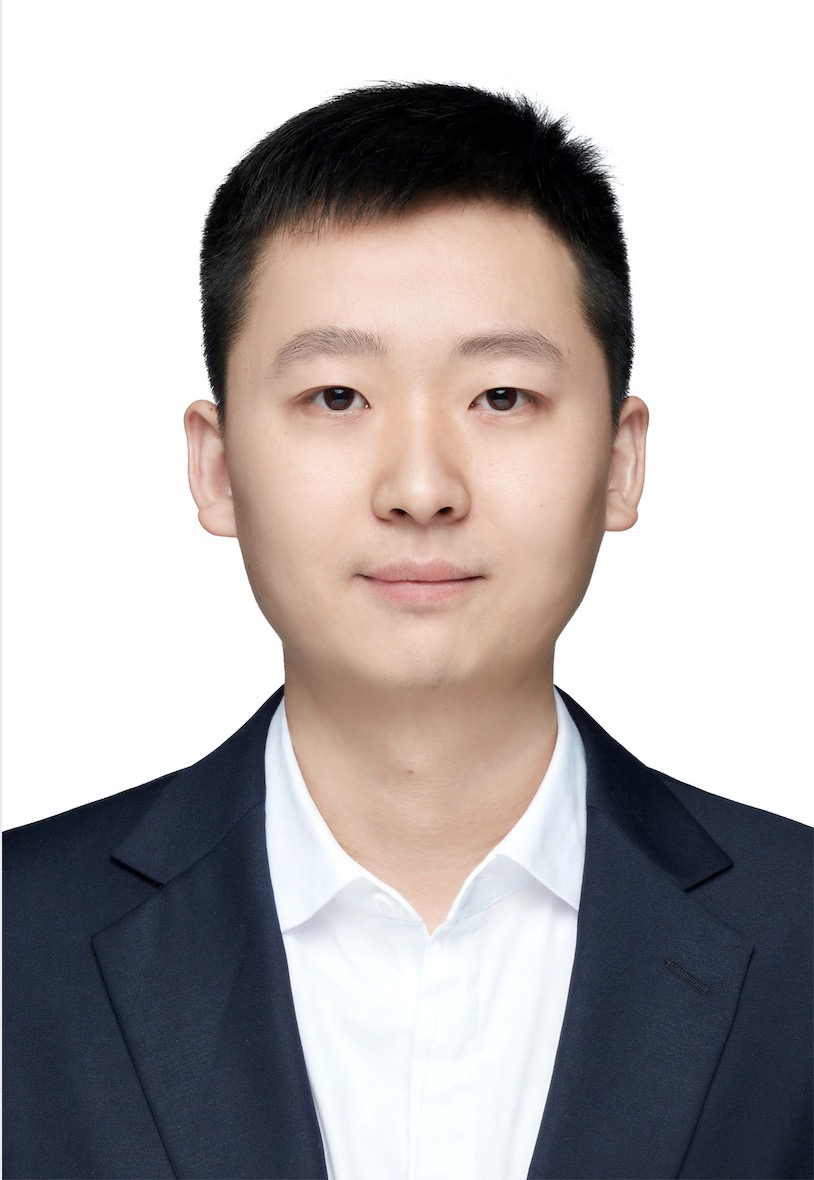}}]{Hao Guo} (Member, IEEE) received the B.S. and M.S. degrees from the Northwest University, Xi'an, China in 2012, and the Illinois Institute of Technology, Chicago, United States in 2014, and his Ph.D. degree from the University of Delaware, Newark, United States in 2020, all in computer science. He is currently an Assistant Professor with the School of Software at the Northwestern Polytechnical University. His research interests include blockchain and distributed ledger technology, data privacy and security, cybersecurity, cryptography technology, and Internet of Things (IoT). He is a member of ACM.
\end{IEEEbiography}

\vskip -6pt plus -1fil
\begin{IEEEbiography}
[{\includegraphics[width=1in,height=1.25in,clip]{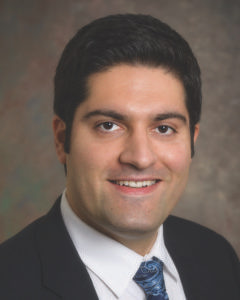}}]{Mark Nejad} (Senior Member, IEEE) is an Assistant Professor at the University of Delaware. His research interests include distributed systems, blockchain, network optimization, and game theory. He has published more than forty peer-reviewed papers and received several publication awards including the 2016 IISE Pritsker best doctoral dissertation award and the 2019 CAVS best paper award from the IEEE VTS. His research is funded by the National Science Foundation and the Department of Transportation. He is a member of INFORMS.
\end{IEEEbiography}

\end{document}